\documentclass[journal]{IEEEtran}

\usepackage{amsfonts}
\usepackage{amsbsy}
\usepackage{amssymb}
\usepackage{amscd}
\usepackage[cmex10]{amsmath}
\usepackage{flushend}
\usepackage{graphicx}
\usepackage[tight,footnotesize]{subfigure}
\usepackage[usenames,dvipsnames]{pstricks}
\usepackage[ruled,vlined,linesnumbered]{algorithm2e}
\usepackage{cite}

\newtheorem{theorem}{Theorem}
\newtheorem{lemma}{Lemma}
\newtheorem{proposition}{Proposition}

\newcommand{\mat}[1]{\boldsymbol{#1}}

\newcommand{\pp}[1]{{\left( #1 \right)}}
\newcommand{\pps}[1]{{( #1 )}}
\newcommand{\ppb}[1]{{\left[ #1 \right]}}
\newcommand{\br}[1]{{\left\{ #1 \right\}}}
\newcommand{\norm}[1]{{ \Vert #1 \Vert }}
\newcommand{\abs}[1]{{ | #1 | }}

\newcommand{\sabs}[1]{{ | #1 |^2 }}
\newcommand{\sabsl}[1]{{ \left| #1 \right|^2 }}
\newcommand{\snorm}[1]{{ \Vert #1 \Vert^2 }}

\newcommand{\gpt}[1]{\check{g}_{#1}}
\newcommand{\g}[2]{{g}_{#1#2}}
\newcommand{\x}[2]{{x}_{#1}^{(#2)}}

\newcommand{\px}[2]{{x}_{#1}^{*(#2)}}
\newcommand{\tx}[2]{{x'}_{#1}^{(#2)}}

\def\mrt{{\text{\tiny{MRT}}}}
\def\core{{\text{\tiny{CORE}}}}
\def\mmse{{\text{\tiny{MMSE}}}}
\def\zf{{\text{\tiny{ZF}}}}
\def\ne{{\text{\tiny{NE}}}}
\def\H{{H}} 
\def\ld{{\log_2}} 
\def\bI{{\mat{I}}} 
\def\bh{{\mat{h}}} 
\def\bw{{\mat{w}}} 
\def\bv{{\mat{v}}} 
\def\bg1{{g_{11}^{\perp}}}
\def\bg2{{g_{22}^{\perp}}}
\def\zfg{{\check{g}}}

\begin{document}

\title{Exchange Economy in Two-User Multiple-Input Single-Output Interference Channels}%
\author{Rami~Mochaourab,~\IEEEmembership{Student~Member,~IEEE,}
        and~Eduard~Jorswieck,~\IEEEmembership{Senior~Member,~IEEE}%
\thanks{\copyright ~2013 IEEE. Personal use of this material is permitted. Permission from IEEE must be obtained for all other uses, in any current or future media, including reprinting/republishing this material for advertising or promotional purposes, creating new collective works, for resale or redistribution to servers or lists, or reuse of any copyrighted component of this work in other works.}
\thanks{Part of this work has been performed in the framework of the European research project SAPHYRE, which is partly funded by the European Union under its FP7 ICT Objective 1.1 - The Network of the Future. This work is also supported in part by the Deutsche Forschungsgemeinschaft (DFG) under grant Jo 801/4-1.}
\thanks{The authors are with the Department of Electrical Engineering and Information Technology, Dresden University of Technology, 01062 Dresden, Germany. E-mail: \{Rami.Mochaourab,Eduard.Jorswieck\}@tu-dresden.de. Phone: +49-351-46332239. Fax: +49-351-46337236.}
\thanks{Part of this work has been presented at IEEE International Conference on Communications, Workshop on Game Theory and Resource Allocation for 4G, Kyoto, Japan, June 5--9, 2011 \cite{Mochaourab2011}.}}

\maketitle

\begin{abstract}
We study the conflict between two links in a multiple-input single-output interference channel. This setting is strictly competitive and can be related to perfectly competitive market models. In such models, general equilibrium theory is used to determine equilibrium measures that are Pareto optimal. First, we consider the links to be consumers that can trade goods within themselves. The goods in our setting correspond to beamforming vectors. We utilize the conflict representation of the consumers in the Edgeworth box, a graphical tool that depicts the allocation of the goods for the two consumers, to provide closed-form solution to all Pareto optimal outcomes. Afterwards, we model the situation between the links as a competitive market which additionally defines prices for the goods. The equilibrium in this economy is called Walrasian and corresponds to the prices that equate the demand to the supply of goods. We calculate the unique Walrasian equilibrium and propose a coordination process that is realized by an arbitrator which distributes the Walrasian prices to the consumers. The consumers then calculate in a decentralized manner their optimal demand corresponding to beamforming vectors that achieve the Walrasian equilibrium. This outcome is Pareto optimal and dominates the noncooperative outcome of the systems. Thus, based on the game theoretic model and solution concept, an algorithm for a distributed implementation of the beamforming problem in multiple-input single-output interference channels is provided.
\end{abstract}

\section{Introduction}
Two transmitter-receiver pairs utilize the same spectral band simultaneously. Each transmitter is equipped with $N$ antennas and each receiver with a single antenna. This setting corresponds to the multiple-input single-output (MISO) interference channel (IFC) \cite{Vishwanath2004}. The systems' performance in such a setting is degraded by mutual interference, and their noncooperative operation is generally not efficient \cite{Larsson2008}. Therefore, coordination between the links is needed in order to improve their joint outcome.

Generally, of interest is to devise coordination mechanisms in which the operating point of the links is Pareto optimal. A Pareto optimal point is an achievable utility tuple from which it is impossible to increase the performance of one link without degrading the performance of another. Consequently, Pareto optimality ensures efficient exploitation of the wireless channel resources. For this purpose, there has been several work on characterizing the set of beamforming vectors that are relevant for Pareto optimal operation in the MISO IFC \cite{Jorswieck2008a, Shang2011, Zhang2010, Mochaourab2011a, Emil2011}. Next, we will discuss these approaches.

\subsection{Characterization of Pareto Optimal Points}

Designing a Pareto optimal mechanism requires finding the joint beamforming vectors used at the transmitters that lead to the Pareto optimal point. The set of feasible beamforming vectors for each transmitter is an $N$-dimensional complex ball where $N$ is the number of used antennas. The importance of characterizing the set of beamforming vectors necessary for the links' Pareto optimal operation is twofold. First, the set of relevant beamforming vectors to consider for finding a Pareto optimal point is reduced to a relatively small subset of all feasible beamforming vectors. Second, the characterized set of efficient beamforming vectors is parameterized by a number of scalars which can even reduce the complexity for indicating the required beamforming vectors.

In \cite{Jorswieck2008a}, the efficient beamforming vectors are parameterized by $K(K-1)$ complex-valued parameters, where $K$ is the number of links. For the special two-user case, the efficient beamforming vectors are proven to be a linear combination of maximum ratio transmission and zero forcing transmission. Thus, two real-valued parameters are required each between zero and one to characterize all Pareto optimal operating points. The extension to a real-valued parametrization for the general $K$-user case is conducted in \cite{Shang2011,Zhang2010,Mochaourab2011a} where $K(K-1)$ real-valued parameters are required to achieve all Pareto optimal points. Recently in \cite{Emil2011}, parametrization of the efficient beamforming vector is provided in the multi-cell MISO setting with general linear transmit power constraint at the transmitters. For the case of MISO IFC and total power constraint at the transmitter, the number of required parameters is $2K - 1$.

In this work, we provide a single real-valued parametrization of the beamforming vectors that are necessary and sufficient to achieve all Pareto optimal points. This result is gained when we model the two-user MISO IFC as a \emph{pure exchange economy} \cite{Jehle2003}. The links are consumers and they possess goods which correspond to beamforming vectors. In a pure exchange economy, the consumers can trade their goods within themselves to improve their utility. The utility function of the consumers in our case is the signal to interference plus noise ratio (SINR) which is formulated in terms of the goods. Utilizing the Edgeworth box \cite{Jehle2003}, which is a graphical tool that depicts the preferences of the consumers over the distribution of the goods, we provide a closed-form solution to all Pareto optimal points of the SINR region. A subset of all Pareto optimal points satisfy that both links jointly achieve higher utility than at the noncooperative point. These points are called \emph{exchange equilibria} and are related to the solution concept, the core, from coalitional game theory.

\subsection{Coordination to Achieve Pareto Optimal Points}

All of the mentioned efforts to parameterize the efficient beamforming vectors in the MISO are valuable for designing efficient low complexity distributed resource allocation schemes such as in \cite{Ho2008,Lindblom2010,Mochaourab2010a}. In \cite{Ho2008} and \cite{Lindblom2010}, the real-valued parametrization for the two-user case from \cite{Jorswieck2008a} is utilized and bargaining algorithms are proposed to improve the joint performance of the systems from the noncooperative state. An extension to these works is made in \cite{Mochaourab2010a} where a strategic bargaining process is proposed and proven to terminate at a Pareto optimal outcome. Also based on a strategic bargaining approach, a coordination mechanism is proposed in \cite{Liu2010} for the two-user MISO IFC where the Han-Kobayashi scheme is applied. In the $K$-user MISO IFC, a low complexity one-shot coordination mechanism is given in \cite{Zakhour2009}, where each transmitter independently maximizes its virtual SINR. For the two-user case, the proposed mechanism is proven to achieve a Pareto optimal solution.

In this paper, we propose a coordination mechanism between two MISO interfering links which is Pareto optimal and achieves for each link a utility higher than at the noncooperation point. Our analysis is based on relating the MISO IFC setting to a competitive market \cite{Jehle2003}. To the best of our knowledge, this is the first time the beamforming problem in the MISO IFC is related to and analyzed using competitive market models. In a competitive market, as proposed by L. Walras \cite{Walras1874,Ye}, there exists a population in which each individual possesses an amount of divisible goods. The worth of these goods makes up the budget of each individual. Each individual has a utility function which reveals his demand on consuming goods. Moreover, each individual would use the revenue from selling all his goods to buy amounts of goods such that his utility is maximized. This economic model is competitive because each consumer seeks to maximize his profit independent of what the other consumers demand. Walras investigated if there exists prices for the goods such that the market has neither shortage nor surplus. The existence of such prices, called Walrasian prices, was later explored by Arrow \cite{Arrow1954}. The prices in this economy are usually assumed to be fixed and not determined by the consumers. It is assumed that the market or an auctioneer acts as an arbitrator to determine the Walrasian prices.

The competitive market model has found a few applications for resource allocation in communication networks. In \cite{Xie2010}, the Walrasian equilibrium is formulated as a linear complementarity problem for a multi-link multi-carrier setting. A decentralized price-adjustment process is proposed where the users send their power allocations in each iteration to the spectrum manager which adjusts the prices to achieve the equilibrium. In \cite{Ye}, competitive spectrum market is considered where the users, sharing a common frequency band, can purchase their transmit power subject to budget constraints. An agent, referred to as the market, determines the unit prices of the power spectra. Existence of the equilibrium is proven and conditions for its uniqueness are provided. In \cite{Tiwari2010}, the competitive equilibrium is used for simultaneous bitrate allocation for multiple video streams and the Edgeworth box \cite{Walras1874} is used to illustrate the conflict between the streams. In the context of cognitive radio, spectrum trading is successfully modeled by economic models and market-equilibrium, and competitive and cooperative pricing schemes are developed in \cite{Niyato2008}. Moreover, in \cite{Niyato2010}, hierarchical spectrum sharing is modeled as an interrelated market. The pricing mechanism for the bandwidth allocations between the systems equates the supply to the demand.

In our case, the links are the consumers and the parameters of the beamforming vectors are the goods the consumers possess. We formulate the consumers' demand functions and calculate the Walrasian prices which equate the demand to the supply of each good. To achieve the Pareto optimal Walrasian equilibrium, the arbitrator coordinates the transmission of the links. We consider two cases for the coordination mechanism. Assuming the arbitrator has full knowledge of the setting, he can calculate the Walrasian prices and forward these to the links. The links independently calculate their beamforming vectors according to their demand function. Assuming the arbitrator has limited knowledge of the setting, we propose a price adjustment process, also referred to as t\^atonnement, to reach the Walrasian prices. In each iteration, the links send their demands to the arbitrator which updates the prices according to the excess demand of each good.

\subsubsection*{Outline}
The outline of the paper is as follows. The system and channel model, as well as the definition of the SINR region and the beamforming vectors that are relevant for Pareto optimal operation are given in Section \ref{sec:Preliminaries}. In Section \ref{sec:exchange}, we examine a pure exchange economy between the links. We model the parametrization of efficient beamforming vectors as goods and the links as consumers which trade these goods within themselves. We characterize all Pareto optimal points in closed form and define the equilibria which correspond to the core of a coalition between the links. In Section \ref{sec:competitive}, we consider a competitive market model and assume that the goods are bought by the consumers at prices determined by an arbitrator. The equilibrium of this market model is determined, and we provide two coordination mechanisms to achieve it. In Section \ref{sec:discussion}, we illustrate the results of this paper before we conclude in Section \ref{sec:conclusion}.

\subsubsection*{Notations}
Column vectors and matrices are given in lowercase and uppercase boldface letters, respectively. $\norm{\mat{a}}$ is the Euclidean norm of $\mat{a} \in \mathbb{C}^{N}$. $\abs{b}$ is the absolute value of $b \in \mathbb{C}$. sign$(a)$ denotes the sign of $a \in \mathbb{R}$. $(\cdot)^T$ and $(\cdot)^\H$ denote transpose and Hermitian transpose, respectively. The orthogonal projector onto the column space of $\mat{Z}$ is $\mat{\Pi}_{Z} := \mat{Z}(\mat{Z}^\H\mat{Z})^{-1}\mat{Z}^\H$. The orthogonal projector onto the orthogonal complement of the column space of $\mat{Z}$ is $\mat{\Pi}_{Z}^{\perp} := \bI - \mat{\Pi}_{Z}$, where $\bI$ is an identity matrix. $\mathcal{CN}(0,\mat{A})$ denotes a circularly-symmetric Gaussian complex random vector with covariance matrix $\mat{A}$. Throughout the paper, the subscripts $k, \ell$ are from the set $\{1,2\}$.%

\section{Preliminaries}\label{sec:Preliminaries}
\subsection{System and Channel Model}
The quasi-static block flat-fading channel vector from transmitter $k$ to receiver $\ell$ is denoted by $\mat{h}_{k\ell} \in \mathbb{C}^{N}$. We assume that transmission consists of scalar coding followed by beamforming. The beamforming vector used by transmitter $k$ is $\mat{w}_{k} \in \mathbb{C}^{N}$. The matched-filtered, symbol-sampled complex baseband data received at receiver $k$ is\footnote{Throughout the paper, the subscripts $k, \ell$ are from the set $\{1,2\}$.}
\begin{equation}
y_k = \bh_{kk}^\H \bw_k s_k + \bh_{\ell k}^\H \bw_\ell s_\ell + n_k, \quad k \neq \ell,
\end{equation}
\noindent where $s_k \sim \mathcal{CN}(0,1)$ is the symbol transmitted by transmitter $k$, and $n_k \sim \mathcal{CN}(0,\sigma^2)$ is additive Gaussian noise. Each transmitter has a total power constraint of $P := 1$ such that $\snorm{\mat{w}_{k}} \leq 1$. We define the signal to noise ratio (SNR) as $1/\sigma^2$.

The transmitters are assumed to have perfect local channel state information (CSI), i.e., each transmitter has perfect knowledge of the channel vectors only between itself and the two receivers. Further information at the transmitters required for the coordination mechanism is discussed later in Section \ref{sec:coordination}.

We assume there exists an \emph{arbitrator} who coordinates the operation of the transmitters. The arbitrator could be any central controller which is connected to both links. Generally, the practical identification of the arbitrator depends on the scenario. For example, in hierarchical networks in which several tiers of networks operate in the same area it is possible that higher network tiers benefit from coordinating the operation of the networks in lower tiers such as in the model used in \cite{Niyato2010}. Moreover, the arbitrator can be the base station of a macrocell which can coordinate the transmission of smaller microcells in its coverage range \cite{Sarnecki1993}. The macrocell base station is usually connected to the microcell base stations via a high capacity link which enables the exchange of channel information required for the coordination process. The applicability of our system model in a cognitive radio network is not suitable if the transmitters are restricted to take into account the interference levels they are allowed to induce at primary receivers. Our setting is suitable for cognitive network settings, in which the users dynamically adapt their transmissions according to the environment these users exist in. A cognitive transmitter can choose with whom it can cooperate and exchange information to improve its utility.

\subsection{SINR Region and Efficient Transmission}%
The signal to interference plus noise ratio (SINR) at receiver $k$ is
\begin{equation}\label{eq:SINR}
\phi_k(\bw_1,\bw_2) = \frac{\sabs{\bh_{kk}^\H \bw_k}}{\sabs{\bh_{\ell k}^\H \bw_\ell} + \sigma^2},\quad k \neq \ell.
\end{equation}
\noindent This results in the achievable rate\footnote{We represent the preference of a link over the used beamforming vectors with the SINR utility function in \eqref{eq:SINR}. The results in this paper also hold for any SINR based utility function which is strictly increasing with SINR such as the achievable rate function.} $\log_2(1 + \phi_k(\bw_1,\bw_2))$ for link $k$ when single user decoding is performed at the receivers. The \emph{SINR region} is the set of all achievable SINR tuples defined as
\begin{equation}\label{eq:RR}
\Phi := \br{\pp{\phi_1(\mat{w}_1,\mat{w}_2),\phi_2(\mat{w}_1,\mat{w}_2)}: \snorm{\mat{w}_{k}} \leq 1}.
\end{equation}%
\noindent In the SINR region, tuples can be ranked according to their Pareto efficiency. An SINR tuple $(\phi'_1,\phi'_2) \in \Phi$ is \emph{Pareto superior} to $(\phi_1,\phi_2) \in \Phi$ if $(\phi'_1,\phi'_2) \geq (\phi_1,\phi_2)$, where the inequality is componentwise and strict for at least one component. The transition from $(\phi_1,\phi_2)$ to $(\phi'_1,\phi'_2)$ is called a \emph{Pareto improvement}. Situations where Pareto improvements are not possible are called \emph{Pareto optimal}. These points constitute the \emph{Pareto boundary} of the SINR region. Formally, the set of Pareto optimal points of $\Phi$ are defined as \cite[p. 18]{Peters1992}
\begin{equation}\label{eq:ParetoBoundary}
\mathcal{P}(\Phi) :=\br{\mat{x} \in \Phi : \text{there is no } \mat{y} \in \Phi \text{ with } \mat{y} \geq \mat{x}, \mat{y} \neq \mat{x}},
\end{equation}
\noindent where the inequality in \eqref{eq:ParetoBoundary} is componentwise.

For the two-user MISO IFC, the set of beamforming vectors that are relevant for Pareto optimal operation are parameterized by a single real-valued parameter $\lambda_k \in [0,1]$ for each transmitter $k\neq \ell$ as \cite[Corollary 1]{Jorswieck2008a}%
\begin{equation}\label{eq:beam_opt_param}
\bw_k(\lambda_k) = \sqrt{\lambda_k} \frac{\mat{\Pi}_{\bh_{k \ell}}\bh_{kk}}{\norm{\mat{\Pi}_{\bh_{k \ell}}\bh_{kk}}} + \sqrt{1-\lambda_k} \frac{ \mat{\Pi}_{\bh_{k \ell}}^\perp \bh_{kk}}{\norm{\mat{\Pi}_{\bh_{k \ell}}^\perp \bh_{kk}}}.
\end{equation}%
\noindent This parametrization is valuable for designing efficient low complexity distributed resource allocation schemes \cite{Mochaourab2010a}. The set of beamforming vector in \eqref{eq:beam_opt_param} includes maximum ratio transmission (MRT) ($\lambda_k^\mrt = \norm{\Pi_{\bh_{k\ell}}\bh_{kk}}^2/\norm{\bh_{kk}}^2$) and zero forcing transmission (ZF) ($\lambda_k^\zf = 0$). According to \cite[Corollary 2]{Jorswieck2008a}, it suffices that the parameters $\lambda_k$ only be from the set $[0,\lambda_k^\mrt]$ for Pareto optimal operation. Note that a transmitter $k$ has to know the channel vectors $\bh_{kk}$ and $\bh_{k\ell}, k \neq \ell,$ perfectly in order to calculate the beamforming vectors in \eqref{eq:beam_opt_param}. Since we are interested in transmissions that lead to Pareto optimal outcomes, we will confine the strategy set of each transmitter to the set in \eqref{eq:beam_opt_param} and formulate the SINR expression in \eqref{eq:SINR} in terms of the parameters $\lambda_k$. For this purpose, we first formulate the power gains at the receivers.
\begin{lemma}\label{lem:powergains}
The power gains at the receivers in terms of the parametrization in \eqref{eq:beam_opt_param} are
\begin{align}\label{eq:gain_dir}
\sabs{\bh_{kk}^\H \bw_k(\lambda_k)} &= (\sqrt{\lambda_k g_{k}} + \sqrt{(1-\lambda_k)\gpt{k}})^2,\\ \label{eq:gain_int}
\sabs{\bh_{k \ell}^\H \bw_k(\lambda_k)} &= \lambda_k \g{k}{\ell}, \quad k \neq \ell,
\end{align}
\noindent where $\lambda_k \in [0,\lambda_k^\mrt]$ and
$g_k := \norm{\mat{\Pi}_{\bh_{k \ell}}\bh_{kk}}^2, ~ \zfg_k := \snorm{\mat{\Pi}^{\perp}_{\bh_{k \ell}} \bh_{kk}}, ~ g_{k\ell} := \snorm{\bh_{k\ell}}$,
\noindent where $k \neq \ell$.
\end{lemma}
\begin{IEEEproof}
The proof is provided in Appendix \ref{proof:powergains}.
\end{IEEEproof}

The SINR of link $k$ can be rewritten using Lemma \ref{lem:powergains} in terms of the parameters in \eqref{eq:beam_opt_param} as
\begin{equation}\label{eq:utility_lambdas}
\phi_k\pps{\lambda_1,\lambda_2} = \frac{\pp{\sqrt{\lambda_k g_{k}} + \sqrt{(1-\lambda_k)\gpt{k}}}^2}{\sigma^2 + \lambda_\ell \g{\ell}{k}}, \quad \ell \neq k.
\end{equation}
\noindent Notice in \eqref{eq:utility_lambdas} that the interference term $\lambda_\ell \g{\ell}{k}$ scales linearly with $\lambda_\ell$. With this respect, the parameter $\lambda_\ell$ can be interpreted as a scaling of interference at the counter receiver. A reduction in $\lambda_\ell$ increases the SINR of link $k$ for fixed $\lambda_k$. Assuming that the links are not cooperative, their operation point can be predicted using noncooperative game theory. The outcome is a solution of a strategic game \cite[Section 2.1]{Osborne1994} between the links.

\subsection{Game in Strategic Form}

In \cite{Larsson2008}, the outcome of a strategic game between the links is studied. The game in strategic form consists of the set of players, $\{1,2\}$, corresponding to the two links. The pure strategies of player $k$ are the real-valued parameters $\lambda_k \in [0,\lambda_k^\mrt]$ in \eqref{eq:beam_opt_param}. The utility function of player $k$ is $\ld(1+\phi_k(\lambda_1,\lambda_2))$, where $\phi_k(\lambda_1,\lambda_2)$ is given in \eqref{eq:utility_lambdas}. The outcome of this strategic game is the same also when the utility function is chosen to be $\phi_k(\lambda_1,\lambda_2)$. This is due to the fact that the preference relation of the players which is represented through the utility function is invariant to positive monotonic transforms \cite[Theorem 1.2]{Jehle2003}. In the above described game, a player always chooses the MRT strategy independent of the choice of the other player \cite{Larsson2008}, i.e., MRT is a dominant strategy for each player. Hence, the unique Nash equilibrium is $(\lambda_1^\mrt,\lambda_2^\mrt)$. The extension of the two-player strategic game described above to the $K$-player case is straightforward. The Nash equilibrium corresponds to the strategy profile in which each player chooses MRT. The outcome in Nash equilibrium is generally not Pareto optimal. In order to achieve Pareto improvements from the Nash equilibrium, coordination between the players is required.

\section{Equilibria in Exchange Economy}\label{sec:exchange}
\subsection{Exchange Economy Model}
In this section, we will use a pure exchange economy model \cite[Chapter 5.1]{Jehle2003} to determine equilibria which lie on the Pareto boundary of the SINR region in \eqref{eq:RR}. This model assumes that there exists a set of consumers which voluntarily exchange goods they possess to increase their payoff. The set of consumers $\{1,2\}$ corresponds to the two links in our setting. The goods correspond to the parametrization of the beamforming vectors in \eqref{eq:beam_opt_param}. That is, there are two goods and $\lambda_1$ will stand for good $1$ and $\lambda_2$ for good $2$. The consumers are initially endowed with amounts of these goods. We will assume that the links start the trade in Nash equilibrium. Thus, consumer $k$ is initially endowed with $\lambda_k^\mrt$ from his good and nothing from the good of the other consumer. Specifically, we define $(\lambda_1^\mrt,0)$ and $(0,\lambda_2^\mrt)$ as the \emph{endowments} of consumers $1$ and $2$, respectively.

Since during exchange each consumer will possess different amounts from both available goods, we introduce new variables that indicate these. When consumer $k$ trades an amount of his good $k$ to consumer $\ell \neq k$, this amount will be represented by $\x{k}{\ell} \leq \lambda_k^\mrt$. The amount left for consumer $k$ from his good is $\x{k}{k} = \lambda_k^\mrt - \x{k}{\ell}$. In connection to the parametrization in \eqref{eq:beam_opt_param}, we define the amounts of possessed goods as
\begin{align}\label{eq:goods1}
\x{k}{k} = \lambda_k, \quad \x{\ell}{k} = \lambda_\ell^\mrt - \lambda_\ell, \quad \ell \neq k.
\end{align}
\noindent If consumer $k$ gives $\x{k}{\ell}$ to the other consumer, this means that transmitter $k$ uses the beamforming vector in \eqref{eq:beam_opt_param} which corresponds to $\lambda_k^\mrt - \x{k}{\ell}$. Hence, if $\x{k}{\ell}$ increases, transmitter $k$ reduces the interference at receiver $\ell$ by using a beamforming vector nearer to ZF. The utility function of a consumer represents his preference over the goods. We use the SINR in \eqref{eq:utility_lambdas} as the utility function of the consumer which we rewrite in terms of the goods as
\begin{equation}\label{eq:utility}
\phi_k\pps{\x1k,\x2k} = \frac{\pp{\sqrt{\x{k}{k} g_{k}} + \sqrt{(1-\x{k}{k})\gpt{k}}}^2}{\sigma^2 + \lambda_\ell^\mrt \g{\ell}{k} - \x{\ell}{k} \g{\ell}{k}},
\end{equation}
\noindent where we substituted $\lambda_k = \x{k}{k}$ and $\lambda_\ell = \lambda_\ell^\mrt - \x{\ell}{k},\ell \neq k,$ from \eqref{eq:goods1}.

\begin{theorem}\label{thm:quasiconcave}
$\phi_k\pps{\x1k,\x2k}$ in \eqref{eq:utility} is continuous, strongly increasing, and strictly quasiconcave on $[0,\lambda_1^\mrt] \times [0,\lambda_2^\mrt]$.
\end{theorem}
\begin{IEEEproof}
The proof is provided in Appendix \ref{proof:quasiconcave}.
\end{IEEEproof}%

\begin{figure}[t]
\centering
\subfigure[Consumer 1.]{
\scalebox{1} 
{
\begin{pspicture}(0.95,-1.7445313)(4.55,1.7245313)
\definecolor{color218b}{rgb}{0.6,0.6,0.6}
\psline[linewidth=0.02cm,linestyle=dashed,dash=0.16cm 0.16cm](1.523125,-0.6354687)(3.103125,-0.6354687)
\pspolygon[linewidth=0.04,linecolor=white,fillstyle=solid,fillcolor=color218b](1.783125,0.94453126)(3.823125,-0.8154687)(3.823125,0.9436222)
\psbezier[linewidth=0.04,fillstyle=solid,fillcolor=color218b](1.863125,0.94453126)(2.023125,0.42453125)(2.423125,-0.73546875)(3.823125,-0.77546877)
\psline[linewidth=0.04cm,arrowsize=0.06cm 3.0,arrowlength=2.0,arrowinset=0.0]{->}(1.423125,-1.2154688)(4.943125,-1.2154688)
\psline[linewidth=0.04cm,arrowsize=0.06cm 3.0,arrowlength=2.0,arrowinset=0.0]{->}(1.583125,-1.3754687)(1.583125,1.7045312)
\usefont{T1}{ptm}{m}{n}
\rput(1.2667187,-1.4804688){\footnotesize $O_1$}
\psline[linewidth=0.02cm](1.503125,0.94453126)(1.663125,0.94453126)
\psline[linewidth=0.02cm](3.823125,-1.3154688)(3.823125,-1.1554687)
\psline[linewidth=0.02cm,linestyle=dashed,dash=0.16cm 0.16cm](3.083125,-1.2954688)(3.083125,-0.6954687)
\usefont{T1}{ptm}{m}{n}
\rput(1.0867188,0.99953127){\footnotesize $\lambda_2^\mrt$}
\usefont{T1}{ptm}{m}{n}
\rput(3.8267188,-1.5404687){\footnotesize $\lambda_1^\mrt$}
\usefont{T1}{ptm}{m}{n}
\rput(0.99671876,1.5395312){\footnotesize $\x21$}
\pscircle[linewidth=0.03,dimen=outer,fillstyle=solid,fillcolor=black](3.093125,-0.6454688){0.07}
\usefont{T1}{ptm}{m}{n}
\rput(4.616719,-1.5404687){\footnotesize $\x11$}
\usefont{T1}{ptm}{m}{n}
\rput(1.2567188,-0.5804688){\footnotesize $\tx21$}
\usefont{T1}{ptm}{m}{n}
\rput(3.1167188,-1.5404687){\footnotesize $\tx11$}
\usefont{T1}{ptm}{m}{n}
\rput(2.5567188,1.1995312){\footnotesize $I_1(\x{1}{1},\phi'_1)$}
\end{pspicture}
}
   \label{fig:subfig1}}
\subfigure[Consumer 2.]{
\scalebox{1} 
{
\begin{pspicture}(1,-1.7345313)(4.6,1.7145313)
\definecolor{color169b}{rgb}{0.6,0.6,0.6}
\psline[linewidth=0.04cm,arrowsize=0.06cm 3.0,arrowlength=2.0,arrowinset=0.0]{->}(1.423125,-1.2254688)(4.943125,-1.2254688)
\psline[linewidth=0.02cm,linestyle=dashed,dash=0.16cm 0.16cm](3.123125,-1.3254688)(3.123125,0.11453125)
\pspolygon[linewidth=0.04,linecolor=white,fillstyle=solid,fillcolor=color169b](1.823125,0.9345313)(3.8039167,-0.74546874)(3.823125,0.9345313)
\psline[linewidth=0.04cm,arrowsize=0.06cm 3.0,arrowlength=2.0,arrowinset=0.0]{->}(1.583125,-1.3854687)(1.583125,1.6945312)
\psbezier[linewidth=0.04,fillstyle=solid,fillcolor=color169b](1.8604984,0.9345313)(2.0473657,0.25453126)(2.7948341,-0.7054688)(3.8039167,-0.7054688)
\usefont{T1}{ptm}{m}{n}
\rput(1.2667187,-1.5104687){\footnotesize $O_2$}
\psline[linewidth=0.02cm](3.783125,-1.3054688)(3.783125,-1.1454687)
\psline[linewidth=0.02cm](1.503125,0.9345313)(1.663125,0.9345313)
\psline[linewidth=0.02cm,linestyle=dashed,dash=0.16cm 0.16cm](1.543125,-0.5654687)(3.163125,-0.5654687)
\usefont{T1}{ptm}{m}{n}
\rput(1.2567188,-0.5104687){\footnotesize $\tx12$}
\usefont{T1}{ptm}{m}{n}
\rput(3.7867188,-1.5304687){\footnotesize $\lambda_2^\mrt$}
\usefont{T1}{ptm}{m}{n}
\rput(4.6367188,-1.5304687){\footnotesize $\x22$}
\usefont{T1}{ptm}{m}{n}
\rput(3.1367188,-1.5504688){\footnotesize $\tx22$}
\usefont{T1}{ptm}{m}{n}
\rput(1.0867188,0.9895313){\footnotesize $\lambda_1^\mrt$}
\usefont{T1}{ptm}{m}{n}
\rput(1.0367187,1.5295312){\footnotesize $\x12$}
\pscircle[linewidth=0.03,dimen=outer,fillstyle=solid,fillcolor=black](3.133125,-0.55546874){0.07}
\usefont{T1}{ptm}{m}{n}
\rput(2.5767188,1.1895312){\footnotesize $I_2(\x{2}{2},\phi'_2)$}
\end{pspicture}
}
   \label{fig:subfig2}}
\caption{\label{fig:edgeworth1}Preference representation of the consumers. $I_1$ and $I_2$ are indifference curves of consumer $1$ and $2$ respectively.}
\end{figure}
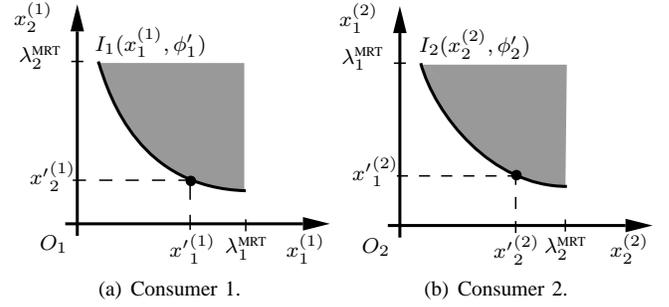

The preference of consumers $1$ and $2$ over the goods is plotted in \figurename~\ref{fig:subfig1} and \figurename~\ref{fig:subfig2}, respectively. For consumer $1$ (analogously consumer $2$), $O_1$ is the origin of the coordinate system which has $\x{1}{1}$, the amount from good $1$, at the x-axis and $\x{2}{1}$, the amount from good $2$, at the y-axis. $I_k$ is the \emph{indifference curve} of consumer $k$ which represents the pairs $(\x1k,\x2k)$ such that the consumer achieves the same payoff as with $(\tx1k,\tx2k)$, i.e., $\phi_k(\x1k,\x2k) = \phi'_k := \phi_k(\tx1k,\tx2k)$. The dark region above $I_k$, corresponds to the pairs $(\x1k,\x2k)$ where the consumer achieves higher payoff than at the indifference curve. The region below $I_k$ corresponds to less payoff for consumer $k$. According to the properties of the utility function in Theorem \ref{thm:quasiconcave}, the indifference curves, which correspond to the boundaries of the level sets of $\phi_k\pps{\x1k,\x2k}$, are convex. This result is required later for the proof of Theorem \ref{thm:demand} to obtain a unique solution to the consumer demand problem in \eqref{prb:demand}. Moreover, Theorem \ref{thm:quasiconcave} proves the existence of at least one Walrasian equilibrium which is considered in Section \ref{sec:competitive}.

Next, we provide two alternative formulations for the indifference curves. Both formulations are required to determine special allocations in the Edgeworth box.

\begin{proposition}\label{thm:ICurve}
The indifference curves $I_k$ ($\x{\ell}{k}$ as a function of $\x{k}{k}$), are calculated for given fixed payoffs $\phi_k'$ as
\begin{align}\label{eq:Itilde1}
I_1(\x11, \phi_1') & = \lambda_2^\mrt + \frac{\sigma^2}{g_{21}} - \frac{\pp{\sqrt{\x11 g_{1}} + \sqrt{(1-\x11)\gpt{1}}}^2}{ \phi'_1 \g{2}{1}}, \\ \label{eq:Itilde2}
I_2(\x22, \phi_2') & = \lambda_1^\mrt + \frac{\sigma^2}{g_{12}} - \frac{\pp{\sqrt{\x22 g_{2}} + \sqrt{(1-\x22)\gpt{2}}}^2}{ \phi'_2 \g{1}{2}}.
\end{align}
\end{proposition}
\begin{IEEEproof}
The indifference curve $I_k$ for a given utility $\phi'_k$ satisfies
\begin{equation}\label{eq:Itilde}
\phi'_k= \frac{\pp{\sqrt{\x{k}{k} g_{k}} + \sqrt{(1-\x{k}{k})\gpt{k}}}^2}{\sigma^2 + \lambda_\ell^\mrt \g{\ell}{k} - \x{\ell}{k} \g{\ell}{k}},\quad \ell \neq k.
\end{equation}
\noindent Exchanging the LHS and the denominator at the RHS of \eqref{eq:Itilde} we get
\begin{equation}
\sigma^2 + \lambda_\ell^\mrt \g{\ell}{k} - \x{\ell}{k} \g{\ell}{k} = \frac{\pp{\sqrt{\x{k}{k} g_{k}} + \sqrt{(1-\x{k}{k})\gpt{k}}}^2}{\phi'_k},
\end{equation}
\noindent Solving for $\x{\ell}{k}$, we get the expressions in \eqref{eq:Itilde1} and \eqref{eq:Itilde2}.
\end{IEEEproof}%
Note that Proposition \ref{thm:ICurve} characterizes a family of indifference curves. Each indifference curve has a domain and range which depends on the fixed SINR value $\phi'_k$. Thus, for selected fixed SINRs, the indifference curves should be restricted to take values in the feasible parameter set from \eqref{eq:beam_opt_param}, i.e., $I_1(\x11, \phi_1') \in [0,\lambda_2^\mrt]$ and $I_2(\x22, \phi_2') \in [0,\lambda_1^\mrt]$. The indifference curves can be alternatively formulated to obtain $\x{k}{k}$ as a function of $\x{\ell}{k}$.
\begin{proposition}\label{thm:ICurveAsilomar}
The indifference curves $\tilde{I}_k$ ($\x{k}{k}$ as a function of $\x{\ell}{k}$), are calculated for given fixed payoffs $\phi_k'$ as \cite[Proposition 1]{Mochaourab2010a}
\begin{align}
\tilde{I}_1(\x21, \phi_1') & = f\pp{\lambda_1^\mrt,\frac{\phi_1'}{\phi_1\pp{\lambda_1^\mrt,\lambda_2^\mrt - \x21}}}, \\
\tilde{I}_2(\x12, \phi_2') & = f\pp{\lambda_2^\mrt,\frac{\phi_2'}{\phi_2\pp{\lambda_1^\mrt - \x12,\lambda_2^\mrt}}},
\end{align}
\noindent where $f(a,b) := (\sqrt{ab} - \sqrt{(1-a)(1-b)})^2$.
\end{proposition}
Similarly, the values of the indifference curves in Proposition \ref{thm:ICurveAsilomar} have to be in the feasible parameter set such that $\tilde{I}_1(\x21, \phi_1')  \in [0,\lambda_1^\mrt]$ and $\tilde{I}_2(\x12, \phi_2')  \in [0,\lambda_2^\mrt]$.
\subsection{Edgeworth Box}\label{sec:edgeworth}

\begin{figure}[t]
\centering
\scalebox{1} 
{
\begin{pspicture}(0.3,-3.02)(8.5,3.02)
\definecolor{color546b}{rgb}{0.6,0.6,0.6}
\psbezier[linewidth=0.04,linecolor=white,fillstyle=solid,fillcolor=color546b](3.0595312,0.9703124)(3.2595313,-0.34968755)(3.7395313,-1.3496876)(6.099531,-1.4296876)
\psbezier[linewidth=0.04,linecolor=white,fillstyle=solid,fillcolor=color546b](3.0195312,0.9703124)(5.4995313,0.81031245)(6.1,-0.78)(6.12,-1.46)
\usefont{T1}{ptm}{m}{n}
\rput(4.8910937,-0.7746875){\footnotesize exchange lens}
\pscircle[linewidth=0.03,dimen=outer,fillstyle=solid,fillcolor=black](6.1295314,-1.4196875){0.07}
\psbezier[linewidth=0.03,linestyle=dashed,dash=0.16cm 0.16cm](1.6195313,-2.0896876)(3.0595312,0.01031245)(6.3795314,0.8503124)(7.2995315,2.0703125)
\usefont{T1}{ptm}{m}{n}
\rput(1.2667187,-2.3746874){\footnotesize $O_1$}
\usefont{T1}{ptm}{m}{n}
\rput(7.6267185,2.3453126){\footnotesize $O_2$}
\psbezier[linewidth=0.04](3.0195312,1.4903125)(3.2195313,-1.2624875)(4.579531,-1.3896875)(7.0195312,-1.4663987)
\psbezier[linewidth=0.04](2.2395313,0.9903125)(5.9870434,0.9903125)(6.102946,-1.2096876)(6.199531,-1.9096875)
\usefont{T1}{ptm}{m}{n}
\rput(2.996719,1.7453125){\footnotesize $I_1$}
\usefont{T1}{ptm}{m}{n}
\rput(1.9767187,1.0653124){\footnotesize $I_2$}
\psline[linewidth=0.0139999995cm,linestyle=dashed,dash=0.16cm 0.16cm](6.119531,-2.1096873)(6.119531,2.0903125)
\psline[linewidth=0.0139999995cm,linestyle=dashed,dash=0.16cm 0.16cm](7.3195314,-1.4096875)(1.6195313,-1.4096875)
\usefont{T1}{ptm}{m}{n}
\rput(7.756719,-1.3746876){\footnotesize $\tx22$}
\usefont{T1}{ptm}{m}{n}
\rput(6.1167192,-2.4146874){\footnotesize $\tx11$}
\usefont{T1}{ptm}{m}{n}
\rput(6.1167192,2.3853126){\footnotesize $\tx12$}
\usefont{T1}{ptm}{m}{n}
\rput(1.2567188,-1.3746876){\footnotesize $\tx21$}
\psbezier[linewidth=0.04](3.3195312,1.4703125)(3.5690768,-0.7096876)(5.8981676,-0.60247445)(6.9795313,-0.60247445)
\psbezier[linewidth=0.04](2.2995312,0.5903124)(4.4425616,0.5903124)(4.957713,-1.1578842)(5.0195312,-1.8896875)
\pscircle[linewidth=0.03,dimen=outer,fillstyle=solid,fillcolor=black](4.0895314,0.0){0.07}
\usefont{T1}{ptm}{m}{n}
\rput(4.7817187,1.5253124){\footnotesize \psframebox*[framesep=0, boxsep=false,fillcolor=white] {contract curve}}
\psbezier[linewidth=0.0139999995](4.794113,1.3303125)(4.739531,0.6682435)(4.284147,0.6546822)(4.330237,0.13031244)
\psbezier[linewidth=0.04](5.8395314,1.8214753)(6.7995315,1.7503124)(6.8195314,1.2903124)(6.9395313,0.81031245)
\psbezier[linewidth=0.04](6.539531,1.9503125)(6.591198,1.3776809)(6.9184203,1.3597862)(7.159531,1.2703124)
\pscircle[linewidth=0.03,dimen=outer,fillstyle=solid,fillcolor=black](6.6895313,1.5003124){0.07}
\psbezier[linewidth=0.04](1.7795312,-0.9845713)(2.6008089,-0.9496875)(2.6753552,-1.4801216)(2.6995313,-1.7496876)
\psbezier[linewidth=0.04](2.0595312,-0.30968755)(2.1795313,-1.1496875)(2.6595314,-1.4296876)(3.3795311,-1.4696876)
\pscircle[linewidth=0.03,dimen=outer,fillstyle=solid,fillcolor=black](2.4495313,-1.1596875){0.07}
\psline[linewidth=0.04cm,arrowsize=0.06cm 3.0,arrowlength=2.0,arrowinset=0.0]{->}(1.4195312,-2.1)(8.199532,-2.1)
\psline[linewidth=0.04cm,arrowsize=0.06cm 3.0,arrowlength=2.0,arrowinset=0.0]{->}(7.4995313,2.1)(0.69953126,2.1)
\psline[linewidth=0.04cm,arrowsize=0.06cm 3.0,arrowlength=2.0,arrowinset=0.0]{->}(1.6195313,-2.3)(1.6195313,3.0)
\psline[linewidth=0.04cm,arrowsize=0.06cm 3.0,arrowlength=2.0,arrowinset=0.0]{->}(7.3195314,2.3)(7.3195314,-3.0)
\usefont{T1}{ptm}{m}{n}
\rput(0.39671874,2.175){\footnotesize $\x12$}
\usefont{T1}{ptm}{m}{n}
\rput(1.1767187,2.755){\footnotesize $\x21$}
\usefont{T1}{ptm}{m}{n}
\rput(8.516719,-2.045){\footnotesize $\x11$}
\usefont{T1}{ptm}{m}{n}
\rput(7.6967187,-2.745){\footnotesize $\x22$}
\end{pspicture}
}
\caption{\label{fig:edgeworth3}An illustration of an Edgeworth box.}
\end{figure}
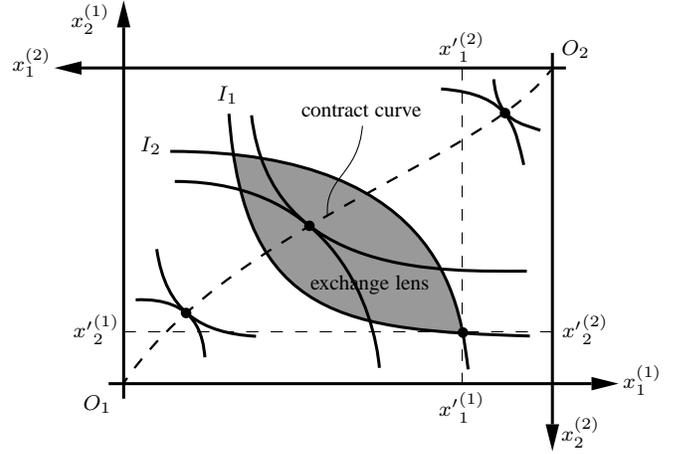

The Edgeworth box \cite{Edgeworth1881}, \cite[Chapter 5]{Jehle2003}, illustrated in \figurename~\ref{fig:edgeworth3}, is a graphical representation that is useful for the analysis of an exchange economy. The box is constructed by joining \figurename~\ref{fig:subfig1} and \figurename~\ref{fig:subfig2}. Thus, the Edgeworth box has two points of origin, $O_1$ and $O_2$, corresponding to consumer $1$ and $2$, respectively. The initial endowments of the consumers define the size of the box. The width of the box is thus $\lambda_1^\mrt$, and the height is $\lambda_2^\mrt$. The possession vectors $(\tx11,\tx21)$ and $(\tx12,\tx22)$ make up the \emph{allocation} $((\tx11,\tx21),(\tx12,\tx22))$ in the box. Every point in the box denotes an allocation, i.e., an assignment of a possession vector to each consumer. The consumers' preferences in the Edgeworth box can be revealed according to their indifference curves. The dark region in \figurename~\ref{fig:edgeworth3} is called the \emph{exchange lens} and contains all allocations that are Pareto improvements to the outcome in $((\tx11,\tx21),(\tx12,\tx22))$. The locus of all Pareto optimal points in the Edgeworth box is called the \emph{contract curve} \cite{Edgeworth1881}. On these points, the indifference curves are tangent, and are characterized by the following condition\footnote{In multiple consumer settings, the condition provided by Edgeworth \cite{Edgeworth1881} should hold for every consumer pair.} \cite[p. 21]{Edgeworth1881}:%
\begin{multline}\label{eq:contractcond}
{\frac{\partial \phi_1\pp{\x11,\x21}}{\partial \x11}}
{\frac{\partial \phi_2\pp{\x12,\x22}}{\partial \x22}}
\\ = {\frac{\partial \phi_2\pp{\x12,\x22}}{\partial \x12}} {\frac{\partial \phi_1\pp{\x11,\x21}}{\partial \x21}}.
\end{multline}%
The convexity of the consumers' indifference curves implies that these can only be tangent at a single point. Thus, the condition in \eqref{eq:contractcond} is necessary and sufficient for an allocation to be on the contract curve.

\begin{theorem}\label{thm:contractcurve}
The contract curve $cc: [0,\lambda_2^\mrt]\rightarrow [0,\lambda_1^\mrt]$ ($\x11$ as a function of $\x22$) is the solution of the following cubic equation\footnote{This result is independently obtained in \cite{Lindblom2011}.}
\begin{align}\label{eq:cubic}
a \ppb{\x11}^3 + b \ppb{\x11}^2 + c \ppb{\x11} + d = 0,
\end{align}
\noindent where
\begin{align}\label{eq:coeffa}
a & = - (g_1 + \zfg_1)(C - g_{12})^2, \quad d = g_1 \sigma^4,\\  \label{eq:coeffb}
b &= (C - g_{12}) \pp{2 \zfg_1(C + \sigma^2) + g_1(2 \sigma^2 + C - g_{12})},\\ \label{eq:coeffc}
c &= -\zfg_1 (C + \sigma^2)^2 + \sigma^2 g_1 (2 g_{12} - 2 C - \sigma^2),
\end{align}
\noindent and $C$ is a function of $\x22$ given as
\begin{equation}
C = \frac{\pp{\sqrt{\x22 g_{2}} + \sqrt{(1-\x22)\gpt{2}}}}{\pp{\sqrt{\frac{g_2}{\x22}} - \sqrt{\frac{\zfg_2}{1-\x22}}} \pp{\frac{\sigma^2}{g_{21}} + \lambda_2^\mrt - \x21}}.
\end{equation}
\noindent The root of interest in \eqref{eq:cubic} lies in $[0,\lambda_1^\mrt]$ and satisfies
\begin{multline}
\text{sign}\pp{{\sigma^2}/{g_{12}} + \x11 - C\x11} \\ = \text{sign}\pp{{\sigma^2}/{g_{12}} + \x11 + C (1-\x11)}.
\end{multline}
\end{theorem}
\begin{IEEEproof}
The proof is provided in \cite[Appendix A]{Mochaourab2011}.
\end{IEEEproof}

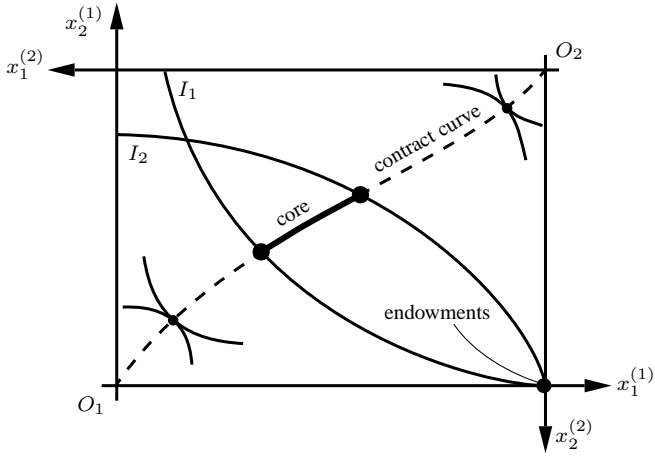
\begin{figure}[t]
\centering
\scalebox{1} 
{
\begin{pspicture}(0.3,-3.02)(8.5,3.02)
\psbezier[linewidth=0.04,linestyle=dashed,dash=0.16cm 0.16cm](1.603125,-2.08)(3.043125,0.02)(6.363125,0.86)(7.283125,2.08)
\psbezier[linewidth=0.08,dotsize=0.07055555cm 2.0]{*-*}(3.523125,-0.32)(4.403125,0.22)(3.963125,-0.04)(4.843125,0.44)
\psbezier[linewidth=0.04](5.943125,1.9111627)(6.903125,1.84)(6.923125,1.38)(7.043125,0.9)
\psbezier[linewidth=0.04](6.643125,2.04)(6.694792,1.4673684)(7.0220137,1.4494737)(7.263125,1.36)
\pscircle[linewidth=0.03,dimen=outer,fillstyle=solid,fillcolor=black](6.793125,1.59){0.07}
\psbezier[linewidth=0.04](1.683125,-1.0548837)(2.5044026,-1.02)(2.578949,-1.5504341)(2.603125,-1.82)
\psbezier[linewidth=0.04](1.963125,-0.38)(2.083125,-1.22)(2.563125,-1.5)(3.283125,-1.54)
\psline[linewidth=0.04cm,arrowsize=0.06cm 3.0,arrowlength=2.0,arrowinset=0.0]{->}(1.403125,-2.1)(8.183125,-2.1)
\psline[linewidth=0.04cm,arrowsize=0.06cm 3.0,arrowlength=2.0,arrowinset=0.0]{->}(7.483125,2.1)(0.683125,2.1)
\psline[linewidth=0.04cm,arrowsize=0.06cm 3.0,arrowlength=2.0,arrowinset=0.0]{->}(1.603125,-2.3)(1.603125,3.0)
\psline[linewidth=0.04cm,arrowsize=0.06cm 3.0,arrowlength=2.0,arrowinset=0.0]{->}(7.303125,2.3)(7.303125,-3.0)
\usefont{T1}{ptm}{m}{n}
\rput(1.2667187,-2.365){\footnotesize $O_1$}
\usefont{T1}{ptm}{m}{n}
\rput(7.6267185,2.355){\footnotesize $O_2$}
\usefont{T1}{ptm}{m}{n}
\rput(2.5567188,1.835){\footnotesize $I_1$}
\usefont{T1}{ptm}{m}{n}
\rput(1.8967187,0.975){\footnotesize $I_2$}
\usefont{T1}{ptm}{m}{n}
\rput{29.314293}(1.319761,-2.6600654){\rput(5.7342186,1.195){\footnotesize contract curve}}
\pscircle[linewidth=0.03,dimen=outer,fillstyle=solid,fillcolor=black](2.353125,-1.23){0.07}
\usefont{T1}{ptm}{m}{n}
\rput(0.39671874,2.175){\footnotesize $\x12$}
\usefont{T1}{ptm}{m}{n}
\rput(1.1767187,2.755){\footnotesize $\x21$}
\usefont{T1}{ptm}{m}{n}
\rput(8.516719,-2.045){\footnotesize $\x11$}
\usefont{T1}{ptm}{m}{n}
\rput(7.6967187,-2.745){\footnotesize $\x22$}
\psbezier[linewidth=0.04](2.243125,2.08)(3.023125,-1.38)(6.643125,-2.12)(7.323125,-2.1)
\psbezier[linewidth=0.04](1.603125,1.24)(5.163125,1.18)(7.143125,-1.2)(7.303125,-2.1)
\usefont{T1}{ptm}{m}{n}
\rput{29.537344}(0.60121167,-1.937418){\rput(3.9642189,0.195){\footnotesize core}}
\psbezier[linewidth=0.0139999995](6.083125,-1.28)(6.343125,-1.66)(6.8106813,-1.9432989)(7.3151884,-2.0935006)
\psdots[dotsize=0.2](7.283125,-2.1)
\usefont{T1}{ptm}{m}{n}
\rput(5.873906,-1.125){\footnotesize endowments}
\end{pspicture}
}
\caption{\label{fig:edgeworth4}An illustration of the allocations in the core.}
\end{figure}

According to Edgeworth \cite{Edgeworth1881}, the outcome of an exchange between the consumers must lie on the contract curve. The solution concept by Edgeworth is related to that of coalitional games called the core \cite{Shubik1961} which defines equilibria in our exchange economy. The situation between the two links can be represented as a coalitional game without transferable payoff \cite[Chapter 13.5]{Osborne1994}. In our case, the core of this game \cite[Definition 268.3]{Osborne1994} is the set of all allocations in the Edgeworth box in which no player can achieve higher payoffs without cooperating with the other player. In \figurename~\ref{fig:edgeworth4}, the core is illustrated as the set of allocations on the contract curve which is bounded by the indifference curves corresponding to the initial endowments. That is, the core allocations correspond to all Pareto optimal points which dominate the Nash equilibrium in the SINR region. With the initial endowments corresponding to the Nash equilibrium $(\lambda_1^\mrt,\lambda_2^\mrt)$, the indifference curves can be calculated from Proposition \ref{thm:ICurveAsilomar} or Proposition \ref{thm:ICurve}. The bounds for the core, as illustrated in \figurename~\ref{fig:edgeworth4}, can be calculated as the intersection points of the indifference curves starting at the endowment allocation and the contract curve characterized in Theorem \ref{thm:contractcurve}. Later in Section \ref{sec:discussion}, the bounds for the core will be used to determine the Kalai-Smorodinsky solution from axiomatic bargaining theory.

\section{Walrasian Equilibrium in Exchange Economy}\label{sec:competitive}

In the preceding section, we have determined the Pareto optimal equilibria in our pure exchange economy. These equilibria can be achieved requiring the links to negotiate or bargain (as for instance is proposed in \cite{Mochaourab2010a}). Next, we will consider decentralized operation of the links and include the arbitrator to coordinate transmission of the links.

\subsection{Competitive Market Model}

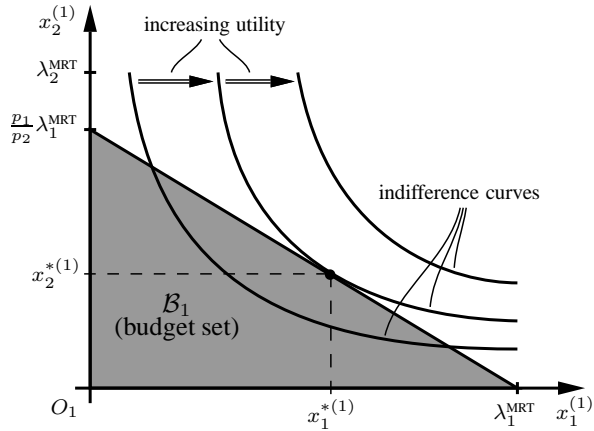
\begin{figure}[t]
\centering
\scalebox{1} 
{
\begin{pspicture}(1.3,-2.8245313)(9.487187,2.8045313)
\definecolor{color80b}{rgb}{0.6,0.6,0.6}
\usefont{T1}{ptm}{m}{n}
\rput(2.3267188,1.9195312){\footnotesize $\lambda_2^\mrt$}
\pspolygon[linewidth=0.04,fillstyle=solid,fillcolor=color80b](2.763125,1.1445312)(2.763125,-2.2954688)(8.443125,-2.2954688)
\psline[linewidth=0.04cm,arrowsize=0.06cm 3.0,arrowlength=2.0,arrowinset=0.0]{->}(2.563125,-2.2954688)(9.343125,-2.2954688)
\psline[linewidth=0.04cm,arrowsize=0.06cm 3.0,arrowlength=2.0,arrowinset=0.0]{->}(2.763125,-2.4954689)(2.763125,2.8045313)
\usefont{T1}{ptm}{m}{n}
\rput(2.4267187,-2.5604687){\footnotesize $O_1$}
\psbezier[linewidth=0.04](3.283125,1.9045312)(3.644325,-0.98362184)(5.563125,-1.7754687)(8.443125,-1.7754687)
\usefont{T1}{ptm}{m}{n}
\rput(8.406719,-2.6404688){\footnotesize $\lambda_1^\mrt$}
\psline[linewidth=0.04cm](2.683125,1.9045312)(2.843125,1.9045312)
\psline[linewidth=0.04cm](8.443125,-2.2154686)(8.443125,-2.3754687)
\usefont{T1}{ptm}{m}{n}
\rput(2.3367188,2.5995312){\footnotesize $\x21$}
\psline[linewidth=0.02cm,linestyle=dashed,dash=0.16cm 0.16cm](5.923125,-0.77546877)(2.683125,-0.77546877)
\psline[linewidth=0.02cm,linestyle=dashed,dash=0.16cm 0.16cm](5.963125,-0.77546877)(5.963125,-2.3754687)
\pscircle[linewidth=0.03,dimen=outer,fillstyle=solid,fillcolor=black](5.953125,-0.78546876){0.07}
\usefont{T1}{ptm}{m}{n}
\rput(9.216719,-2.6004686){\footnotesize $\x11$}
\usefont{T1}{ptm}{m}{n}
\rput(3.9145312,-1.1854688){$\mathcal{B}_1$}
\usefont{T1}{ptm}{m}{n}
\rput(2.1367188,1.1595312){\footnotesize $\frac{p_1}{p_2}\lambda_1^\mrt$}
\psline[linewidth=0.04cm](2.683125,1.1445312)(2.843125,1.1445312)
\psbezier[linewidth=0.04](5.523125,1.9045312)(5.9286804,-0.4646995)(7.6725693,-0.8954688)(8.443125,-0.8954688)
\usefont{T1}{ptm}{m}{n}
\rput(2.3167188,-0.7604687){\footnotesize $\px21$}
\usefont{T1}{ptm}{m}{n}
\rput(5.976719,-2.6404688){\footnotesize $\px11$}
\psbezier[linewidth=0.04](4.463125,1.9045312)(4.683125,-1.0154687)(7.243125,-1.3754687)(8.443125,-1.4014688)
\psline[linewidth=0.02cm,arrowsize=0.04cm 2.0,arrowlength=2.0,arrowinset=0.0,doubleline=true,doublesep=0.02,doublecolor=white]{->}(3.403125,1.7845312)(4.403125,1.7845312)
\usefont{T1}{ptm}{m}{n}
\rput(4.4275,2.5195312){\footnotesize increasing utility}
\psline[linewidth=0.02cm,arrowsize=0.04cm 2.0,arrowlength=2.0,arrowinset=0.0,doubleline=true,doublesep=0.02,doublecolor=white]{->}(4.563125,1.7845312)(5.483125,1.7845312)
\psbezier[linewidth=0.02](4.243125,2.3445313)(4.163125,2.1845312)(3.883125,2.0645313)(3.843125,1.9045312)
\psbezier[linewidth=0.02](4.483125,2.3445313)(4.563125,2.1845312)(4.843125,2.0645313)(4.883125,1.9045312)
\usefont{T1}{ptm}{m}{n}
\rput(7.675156,0.31953126){\footnotesize indifference curves}
\usefont{T1}{ptm}{m}{n}
\rput(3.9182813,-1.5054687){(budget set)}
\psbezier[linewidth=0.02](7.683125,0.10453125)(7.539125,-0.55318546)(6.815125,-1.0177631)(6.683125,-1.5954688)
\psbezier[linewidth=0.02](7.723125,0.10453125)(7.635125,-0.37546876)(7.327125,-0.73546875)(7.283125,-1.2154688)
\psbezier[linewidth=0.02](7.763125,0.10453125)(7.715125,-0.2009233)(7.627125,-0.3900142)(7.603125,-0.6954687)
\end{pspicture}
}
\caption{\label{fig:Budgetset}An illustration of the budget set of consumer $1$.}
\end{figure}

In a competitive market, the consumers buy quantities of goods and also sell goods they possess such that they maximize their profit. Each good has a price and every consumer takes the prices as given. The prices of the goods are not determined by consumers, but arbitrated by markets. In our case, the arbitrator determines the prices of the goods. Let $p_k$ denote the unit price of good $k$. In order to be able to buy goods, each consumer $k$ is endowed with a budget $\lambda_k^\mrt p_k$ which is the worth of his initial amounts of goods\footnote{This case corresponds to the Arrow-Debreu market model \cite{Ye}.}. The \emph{budget set} of consumer $k$ is the set of bundles of goods he can afford to possess defined as
\begin{equation}\label{eq:budgetset}
\mathcal{B}_k := \br{\pps{\x1k,\x2k} \in \mathbb{R}_+^2:\x1k p_1 + \x2k p_2 \leq \lambda_k^\mrt p_k}.
\end{equation}
\noindent The budget set of consumer $1$ is illustrated by the grey area in \figurename~\ref{fig:Budgetset}. The boundary of the budget set is a line which connects the points $(\lambda_1^\mrt,0)$ and $(0, \lambda_1^\mrt p_1/p_2)$. Thus, the boundary has a slope of $-p_1/p_2$. For the consumers, the prices of the goods are measures for their qualitative valuation. If $p_1$ is greater than $p_2$, then good $1$ has more value than good $2$. Given the prices $p_1$ and $p_2$, consumer $1$ demands the amounts of goods $\x11$ and $\x21$ such that these maximize his utility in \eqref{eq:utility}. Thus, consumer $k$ solves the following problem:
\begin{equation}\label{prb:demand}
\begin{split}
\text{maximize } & \quad \phi_k\pp{\x1k,\x2k}\\
\text{subject to } & \quad p_1 \x1k + p_2 \x2k \leq \lambda_k^\mrt p_k .
\end{split}
\end{equation}
\noindent In the above consumer problem, the objective function is the SINR of link $k$ and the constraint is defined by the budget set of consumer $k$ in \eqref{eq:budgetset}. The physical interpretation of the budget set constraint can be related to an interference constraint. Considering consumer $1$, the constraint in \eqref{prb:demand} can be reformulated to
\begin{equation}\label{eq:intconstraint}
\x{1}{1} \leq \lambda_1^\mrt  - \frac{p_2}{p_1} \x{2}{1},
\end{equation}
\noindent where, as mentioned before, $\x{1}{1} = \lambda_1 \in [0,\lambda_1^\mrt]$ is the scaling of interference transmitter $1$ produces at receiver $2$. Analogously, $\x{2}{1} = \lambda_2^\mrt - \lambda_2$ is the scaling for interference reduction from transmitter $2$ at receiver $1$. Hence, the constraint in \eqref{eq:intconstraint} dictates the tradeoff between the amount of interference transmitter $1$ can generate at receiver $2$ and the amount of interference receiver $1$ is to tolerate. The prices $p_1$ and $p_2$ can be interpreted as parameters to control the fairness between the links by regulating the amount of interference the links generate on each other.

\begin{theorem}\label{thm:demand}
The unique solution to the problem in \eqref{prb:demand} is
\begin{align}\label{eq:demand1}
\px11(p_1,p_2) & = \frac{1}{1 + \frac{\zfg_1}{g_1}\pp{1 + \frac{g_{21} \frac{p_1}{p_2}}{{\sigma^2} + \lambda_2^\mrt g_{21} - \lambda_1^\mrt g_{21} \frac{p_1}{p_2}}}^2  },\\ \label{eq:demand11}
\px21(p_1,p_2) & = \frac{p_1}{p_2} \pp{\lambda_1^\mrt  - \px11},
\end{align}
\noindent for consumer $1$, and
\begin{align}\label{eq:demand2}
\px22(p_1,p_2) & = \frac{1}{1 + \frac{\zfg_2}{g_2}\pp{1 + \frac{g_{12}\frac{p_2}{p_1}}{{\sigma^2} + \lambda_1^\mrt g_{12} - \lambda_2^\mrt g_{12}\frac{p_2}{p_1}}}^2},\\ \label{eq:demand22}
\px12(p_1,p_2) & = \frac{p_2}{p_1} \pp{\lambda_2^\mrt  - \px22},
\end{align}
\noindent for consumer $2$, where $\zfg_k, g_k, g_{k\ell}$ are defined in Lemma \ref{lem:powergains}. The feasible prices ratio are in the range:
\begin{equation}\label{eq:condPrices}
\underline{\beta}:=\frac{{\lambda_2^\mrt} g_{12}}{{{\sigma^2} + \lambda_1^\mrt g_{12}}} \leq \frac{p_1}{p_2} \leq \overline{\beta} := \frac{{\sigma^2} + \lambda_2^\mrt {g_{21}}}{\lambda_1^\mrt {g_{21}}}.
\end{equation}
\end{theorem}
\begin{IEEEproof}
The proof is provided in Appendix \ref{proof:demand}.
\end{IEEEproof}

Theorem \ref{thm:demand} characterizes the demand functions of each consumer. In economic theory, these functions are called \emph{Marshallian demand functions} \cite{Jehle2003} or \emph{Walrasian demand functions} \cite{Mas-Colell1995}. Note that each consumer calculates his demands independently without knowing the other consumer's demands. From Theorem \ref{thm:demand}, consumer 1 (analogously consumer 2) needs to know the constants $g_1, \zfg_1,$ and $g_{21}$. The measure ${\sigma^2} + \lambda_2^\mrt g_{21}$ in \eqref{eq:utility} is the noise plus interference power in Nash equilibrium. This measure is reported from receiver $1$ to its transmitter at Nash equilibrium which is the initial state of the links before coordination takes place.

The demand functions of the consumers in Theorem \ref{thm:demand} are homogenous of degree zero \cite[Definition A2.2]{Jehle2003} with the prices $p_1$ and $p_2$. That is, the demand of consumer $1$ for good $1$ (analogously consumer $2$ for good $2$) satisfies $\px11(t p_1, t p_2) = \px11(p_1,p_2)$ for $t > 0$. Hence, given only a prices ratio $\bar{p}_1/\bar{p}_2$, we can calculate a prices pair as $p_1 = \bar{p}_1/\bar{p}_2$ and $p_2 = 1$ which leads to the same demand as with $\bar{p}_1$ and $\bar{p}_2$. With this respect, a consumer need only know the price ratio $p_1/p_2$ from the arbitrator to calculate his demands. In \figurename~\ref{fig:Budgetset}, the demand of consumer $1$ is illustrated as the point where the corresponding indifference curve is tangent to the boundary of the budget set.

The next result provides a significant property that the goods in our setting possess. Later in Section \ref{sec:Walras} and Section \ref{sec:coordination}, this property is required to prove the uniqueness of the Walrasian equilibrium and also to guarantee the global convergence of the price adjustment process.

\begin{lemma}\label{thm:uniqueness}
The goods in our setting are \emph{gross substitutes}, i.e., increasing the price of one good increases the demand of the other good.
\end{lemma}
\begin{IEEEproof}
Decreasing the ratio $p_1/p_2$ can be interpreted as decreasing $p_1$ or increasing $p_2$. Consider the \emph{aggregate excess demand} of good $1$ defined as
\begin{equation}\label{eq:excessDemand}
    z_1(p_1,p_2) = \px11(p_1,p_2) + \px12(p_1,p_2) - \lambda_1^\mrt,
\end{equation}
\noindent where $\px{1}{1}(p_1,p_2)$ and $\px{1}{2}(p_1,p_2)$ are the demand functions of good $1$ in \eqref{eq:demand1} and \eqref{eq:demand22} from Theorem \ref{thm:demand}. If $p_1/p_2$ decreases, then $\px11(p_1,p_2)$ increases. If $p_1/p_2$ decreases, then $\px12(p_1,p_2)$ also increases since $p_2/p_1$ increases and $\px22(p_1,p_2)$ decreases. Thus, the aggregate excess demand of good $1$ in \eqref{eq:excessDemand} increases if $p_1/p_2$ decreases. The analysis is analogous for the second good.
\end{IEEEproof}

If each consumer is to demand amounts of goods without considering the demands of the other consumer, then it is important that the consumers' demands equal the consumers' supply of goods. Prices which fulfill this requirement are called \emph{Walrasian} and are calculated next.

\subsection{Walrasian Equilibrium}\label{sec:Walras}
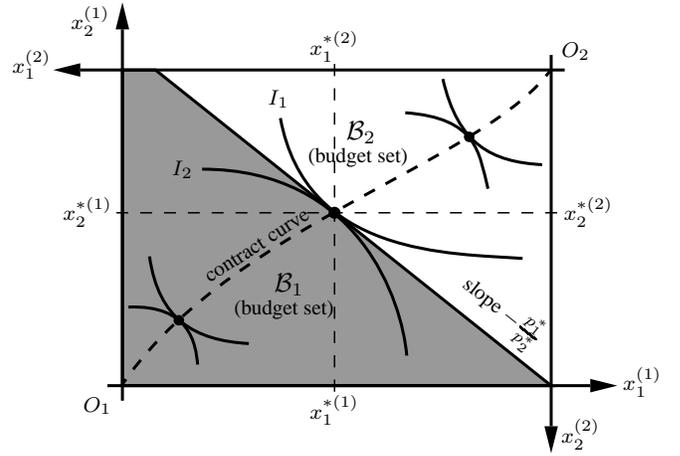
\begin{figure}[t]
\centering
\scalebox{1} 
{
\begin{pspicture}(0.3,-3.02)(8.5,3.02)
\definecolor{color2289b}{rgb}{0.6,0.6,0.6}
\pspolygon[linewidth=0.04,fillstyle=solid,fillcolor=color2289b](2.043125,2.1)(1.603125,2.1)(1.603125,-2.1)(7.303125,-2.1)
\psline[linewidth=0.02cm,linestyle=dashed,dash=0.16cm 0.16cm](4.423125,2.18)(4.423125,-2.18)
\psline[linewidth=0.02cm,linestyle=dashed,dash=0.16cm 0.16cm](1.523125,0.2)(7.383125,0.2)
\psbezier[linewidth=0.04](5.363125,1.5311627)(6.323125,1.46)(6.343125,1.0)(6.463125,0.52)
\psbezier[linewidth=0.04](5.863125,1.98)(6.0639944,1.1098751)(6.508777,0.87680596)(7.183125,0.84)
\psbezier[linewidth=0.04](1.683125,-1.0548837)(2.5044026,-1.02)(2.578949,-1.5504341)(2.603125,-1.82)
\psbezier[linewidth=0.04](1.963125,-0.38)(2.083125,-1.22)(2.563125,-1.5)(3.283125,-1.54)
\psbezier[linewidth=0.04,linestyle=dashed,dash=0.16cm 0.16cm](1.603125,-2.08)(3.043125,0.02)(6.363125,0.86)(7.283125,2.08)
\psline[linewidth=0.04cm,arrowsize=0.06cm 3.0,arrowlength=2.0,arrowinset=0.0]{->}(1.403125,-2.1)(8.183125,-2.1)
\psline[linewidth=0.04cm,arrowsize=0.06cm 3.0,arrowlength=2.0,arrowinset=0.0]{->}(7.483125,2.1)(0.683125,2.1)
\psline[linewidth=0.04cm,arrowsize=0.06cm 3.0,arrowlength=2.0,arrowinset=0.0]{->}(1.603125,-2.3)(1.603125,3.0)
\psline[linewidth=0.04cm,arrowsize=0.06cm 3.0,arrowlength=2.0,arrowinset=0.0]{->}(7.303125,2.3)(7.303125,-3.0)
\usefont{T1}{ptm}{m}{n}
\rput(1.2667187,-2.365){\footnotesize $O_1$}
\usefont{T1}{ptm}{m}{n}
\rput(7.6267185,2.355){\footnotesize $O_2$}
\usefont{T1}{ptm}{m}{n}
\rput(3.6767187,1.715){\footnotesize $I_1$}
\usefont{T1}{ptm}{m}{n}
\rput(2.3967187,0.775){\footnotesize $I_2$}
\psbezier[linewidth=0.04](2.663125,0.78)(4.743125,0.78)(5.3213067,-0.96819675)(5.383125,-1.7)
\pscircle[linewidth=0.03,dimen=outer,fillstyle=solid,fillcolor=black](4.423125,0.2){0.08}
\usefont{T1}{ptm}{m}{n}
\rput{32.85952}(0.4216305,-1.8838977){\rput(3.3942187,-0.225){\footnotesize contract curve}}
\pscircle[linewidth=0.03,dimen=outer,fillstyle=solid,fillcolor=black](6.213125,1.21){0.07}
\pscircle[linewidth=0.03,dimen=outer,fillstyle=solid,fillcolor=black](2.353125,-1.23){0.07}
\usefont{T1}{ptm}{m}{n}
\rput{-38.684044}(2.171965,3.9082315){\rput(6.637344,-1.125){\footnotesize slope $-\frac{p^*_1}{p^*_2}$}}
\usefont{T1}{ptm}{m}{n}
\rput(3.8145313,-0.75){$\mathcal{B}_1$}
\usefont{T1}{ptm}{m}{n}
\rput(4.7745314,1.27){$\mathcal{B}_2$}
\psbezier[linewidth=0.04](3.703125,1.46)(4.043125,-0.38)(5.963125,-0.34)(6.923125,-0.40737557)
\usefont{T1}{ptm}{m}{n}
\rput(3.7532814,-1.085){\footnotesize (budget set)}
\usefont{T1}{ptm}{m}{n}
\rput(4.733281,0.935){\footnotesize (budget set)}
\usefont{T1}{ptm}{m}{n}
\rput(4.416719,-2.425){\footnotesize $\px11$}
\usefont{T1}{ptm}{m}{n}
\rput(1.1367188,0.215){\footnotesize $\px21$}
\usefont{T1}{ptm}{m}{n}
\rput(4.416719,2.415){\footnotesize $\px12$}
\usefont{T1}{ptm}{m}{n}
\rput(7.7967186,0.215){\footnotesize $\px22$}
\usefont{T1}{ptm}{m}{n}
\rput(0.39671874,2.175){\footnotesize $\x12$}
\usefont{T1}{ptm}{m}{n}
\rput(1.1767187,2.755){\footnotesize $\x21$}
\usefont{T1}{ptm}{m}{n}
\rput(8.516719,-2.045){\footnotesize $\x11$}
\usefont{T1}{ptm}{m}{n}
\rput(7.6967187,-2.745){\footnotesize $\x22$}
\end{pspicture}
}
\caption{\label{fig:EdgeworthBox_exp}An illustration of an Edgeworth box. $I_1$ and $I_2$ are indifference curves of consumer $1$ and $2$ respectively. The line with slope -$p_1^*/p^*_2$ separates the budget sets of the consumers in Walrasian equilibrium.}
\end{figure}

In a Walrasian equilibrium, the demand equals the supply of each good \cite[Definition 5.5]{Jehle2003}. According to the properties of the utility function in Theorem \ref{thm:quasiconcave}, there exists at least one Walrasian equilibrium \cite[Theorem 5.5]{Jehle2003}. The Walrasian prices $(p^*_1,p^*_2)$ that lead to a Walrasian equilibrium satisfy
\begin{align}\label{eq:demsup1}
\px11(p_1,p_2) + \px12(p_1,p_2) & = \lambda_1^\mrt, \\ \label{eq:demsup2} \text{and} \quad
\px21(p_1,p_2) + \px22(p_1,p_2) & = \lambda_2^\mrt.
\end{align}
\noindent In our setting in which only two goods exist, Walras' law \cite[Chapter 5.2]{Jehle2003} provides the property that if the demand equals the supply of one good, then the demand would equal the supply of the other good. Hence, in order to calculate the Walrasian prices, it is sufficient to consider only one of the conditions in \eqref{eq:demsup1} and \eqref{eq:demsup2}.
\begin{theorem}\label{thm:Walras}
The ratio of the Walrasian prices is the unique root of
\begin{equation}\label{eq:Wroots}
a \ppb{\frac{p_1}{p_2}}^5 + b \ppb{\frac{p_1}{p_2}}^4 + c \ppb{\frac{p_1}{p_2}}^3 + d \ppb{\frac{p_1}{p_2}}^2 + e \ppb{\frac{p_1}{p_2}} +f = 0,
\end{equation}
\noindent that satisfies the condition in \eqref{eq:condPrices}. The constant coefficients are
\begin{align} \nonumber
a &= T_1 T_2^2 T3, \quad b =-2T_3T_2(T_2S_2+T_1S_1), \\ \nonumber
c &=2T_4T_2S_3+4S_1S_2T_2T_3+T_1S_4T_3,\\ \nonumber
d &=-2S_4S_2T_3 - 4T_1T_2S_2S_3 - S_1T_4S_3,\\ \nonumber
e &=2S_3S_2(T_2S_2+T_1S_1), \quad f = - S_1 S_2^2 S_3,
\end{align}
\noindent where
\begin{align}\nonumber
T_1 &= \pp{g_1 - \zfg_1}/\pp{g_1 + \zfg_1}, \quad T_2 = \lambda_1^\mrt + {\sigma^2}/{g_{12}},\\ \nonumber
T_3 &= (1-\lambda_1^\mrt)\lambda_1^\mrt, \quad T_4 = \pp{\zfg_1^2 - \zfg_1 g_1 + g_1^2}/{(g_1 + \zfg_1)^2},\\ \nonumber
S_1 &= \pp{g_2 - \zfg_2}/\pp{g_2 + \zfg_2}, \quad S_2 = \lambda_2^\mrt + {\sigma^2}/{g_{21}},\\ \nonumber
S_3 &= (1-\lambda_2^\mrt)\lambda_2^\mrt, \quad S_4 = \pp{\zfg_2^2 - \zfg_2 g_2 + g_2^2}/{(g_2 + \zfg_2)^2},
\end{align}
\noindent and $\zfg_k, g_k, g_{k\ell}$ are defined in Lemma \ref{lem:powergains}.
\end{theorem}
\begin{IEEEproof}
Substituting \eqref{eq:demand1} and \eqref{eq:demand22} in \eqref{eq:demsup1} and collecting $p_1/p_2$ we get the expression in \eqref{eq:Wroots}. The condition in \eqref{eq:condPrices} states the set of feasible prices such that the demands of the consumers are feasible. At least one price pair is in this set since a Walrasian equilibrium always exists in our setting. In addition, having the property that the goods are gross substitutes in Lemma \ref{thm:uniqueness}, implies that the Walrasian equilibrium in our setting is unique \cite[Proposition 17.F.3]{Mas-Colell1995}.
Note that the roots in \eqref{eq:Wroots} can be easily calculated using a Newton method. And due to the uniqueness of the Walrasian prices, only one root satisfies the condition in \eqref{eq:condPrices}.
\end{IEEEproof}

According to the First Welfare Theorem \cite[Theorem 5.7]{Jehle2003}, the Walrasian equilibrium is Pareto optimal. Moreover, linking to the results in the previous section, the Walrasian equilibrium lies in the core \cite[Theorem 5.6]{Jehle2003}. In other words, the Walrasian equilibrium dominates the Nash equilibrium outcome. In \figurename~\ref{fig:EdgeworthBox_exp}, the allocation in Walrasian equilibrium which corresponds to the Walrasian prices ratio $p_1^*/p^*_2$ is illustrated in the Edgeworth box. It is the point on the contract curve which intersects the line that passes through the endowment point (Nash equilibrium) with slope $-p_1^*/p^*_2$ (with respect to the coordinate system of consumer $1$). The grey area in \figurename~\ref{fig:EdgeworthBox_exp} is the budget set of consumer $1$ as described in \figurename~\ref{fig:Budgetset}. The white area in the Edgeworth box is the budget set of consumer $2$. According to the axis transformation in constructing the Edgeworth box, the boundaries of the consumers' budget sets coincide. The indifference curves of the consumers are tangent to this line and also tangent to one another which illustrates the Pareto optimality of the Walrasian equilibrium.

\subsection{Coordination Mechanism}\label{sec:coordination}

\begin{table}[t]
  \centering
  \caption{Required information at the arbitrator and transmitters to implement the Walrassian equilibrium in one-shot.}\label{tbl:CSI1}
  \begin{tabular}{r|l}
     & Information \\
     \hline
    Arbitrator & $\bh_{11}, \bh_{12}, \bh_{21}, \bh_{22}, \sigma^2$ \\
    Transmitter $1$ & $\bh_{11}, \bh_{12}, \sigma^2 + \lambda_2^\mrt \norm{\bh_{21}}^2, \norm{\bh_{21}}^2$  \\
    Transmitter $2$ & $\bh_{22}, \bh_{21}, \sigma^2 + \lambda_1^\mrt \norm{\bh_{12}}^2, \norm{\bh_{12}}^2$
  \end{tabular}
\end{table}

In this section, we provide two coordination mechanisms which require different amount of information at the arbitrator. If the arbitrator has full knowledge of all parameters of the setting, then he can calculate the Walrasian prices from Theorem \ref{thm:Walras} and forward these to the transmitters. The transmitters calculate their demands from Theorem \ref{thm:demand} and choose the beamforming vectors accordingly. This mechanism that uses the results in Theorem \ref{thm:demand} and Theorem \ref{thm:Walras} leads directly to the Walrasian equilibrium. In Table \ref{tbl:CSI1}, the required information at the arbitrator and the transmitters to implement this one-shot mechanism are listed. We assume that each transmitter forwards the channel information it has to the arbitrator. Note that each transmitter $k$ initially knows the channel vectors $\bh_{kk}$ and $\bh_{k\ell}, k \neq \ell$, which are required to calculate the efficient beamforming vectors in \eqref{eq:beam_opt_param}. Also, transmitter $k$ knows the sum $\sigma^2 + \lambda_\ell^\mrt \norm{\bh_{\ell k}}^2, k \neq \ell,$ since this is the noise plus interference in Nash equilibrium forwarded through feedback from the intended receiver. The arbitrator, which now has full knowledge of all channels, can then forward the missing information on the channel gain $\norm{\bh_{\ell k}}^2$ to a transmitter $k$ .

If the arbitrator has limited information about the setting, we could still achieve the Walrasian prices through an iterative price adjustment process. For fixed arbitrary initial prices, the transmitters can calculate their demands and forward these to the arbitrator. The arbitrator exploits the demand information to update the prices of the goods. Specifically, the arbitrator would increase the price of the good which has higher demand than its supply. Due to the properties of the goods in Lemma \ref{thm:uniqueness}, this price adjustment process, also called t\^atonnement, is globally convergent to the Walrasian prices given in Theorem \ref{thm:Walras} \cite{Arrow1959}. The price adjustment process requires the information listed in Table \ref{tbl:CSI2} to be available at the arbitrator and the transmitters. In contrast to Table \ref{tbl:CSI1}, the arbitrator requires aside from the noise power $\sigma^2$ only the cross channel gains $\norm{\bh_{21}}^2, \norm{\bh_{12}}^2$ and the parameters $\lambda_1^\mrt,\lambda_2^\mrt$ from the transmitters. This information is required only at the beginning of the price adjustment process in order to calculate the bounds for the feasible prices $\underline{\beta}$ and $\overline{\beta}$ given in \eqref{eq:condPrices}.
\begin{table}[t]
  \centering
  \caption{Required information at the arbitrator and transmitters for the price adjustment process.}\label{tbl:CSI2}
  \begin{tabular}{r|l}
     & Information \\
     \hline
    Arbitrator & $\norm{\bh_{21}}^2, \norm{\bh_{12}}^2,\lambda_1^\mrt,\lambda_2^\mrt, \sigma^2$ \\
    Transmitter $1$ & $\bh_{11}, \bh_{12}, \sigma^2 + \lambda_2^\mrt \norm{\bh_{21}}^2, \norm{\bh_{21}}^2$  \\
    Transmitter $2$ & $\bh_{22}, \bh_{21}, \sigma^2 + \lambda_1^\mrt \norm{\bh_{12}}^2, \norm{\bh_{12}}^2$
  \end{tabular}
\end{table}

\begin{algorithm}[h]
    \IncMargin{2em}
    \KwIn{$\x11,\x12,\x21,\x22$}
    initialize: accuracy $\epsilon$, $n = 0$, $\overline{\beta}^{(0)} = \overline{\beta}$, $\underline{\beta}^{(0)} = \underline{\beta}$ in \eqref{eq:condPrices}, $\frac{p_1^{(0)}}{p_2^{(0)}} = \frac{\overline{\beta}^{(0)}}{2} + \frac{\underline{\beta}^{(0)}}{2}$\;
    \While {$ {\overline{\beta}^{(n)} - \underline{\beta}^{(n)}} > \epsilon$}
    {
    receive demands $\x11,\x12,\x21,\x22$\;
    $n = n +1$\;
    \eIf {$\x11 + \x12 > \lambda_1^\mrt$}{
     $\underline{\beta}^{(n)} = \frac{p_1^{(n-1)}}{p_2^{(n-1)}}$, $\overline{\beta}^{(n)} = \overline{\beta}^{(n-1)}$\;
     $\frac{p_1^{(n)}}{p_2^{(n)}} = \frac{{\overline{\beta}^{(n)}} + {\underline{\beta}^{(n)}}}{2}$\;
    }{
     $\underline{\beta}^{(n)} = \underline{\beta}^{(n-1)}$, $\overline{\beta}^{(n)} = \frac{p_1^{(n-1)}}{p_2^{(n-1)}}$\;
     $\frac{p_1^{(n)}}{p_2^{(n)}} = \frac{{\overline{\beta}^{(n)}} + {\underline{\beta}^{(n)}}}{2}$\;
    }
    }
    \KwOut{${p_1^{(n)}}/{p_2^{(n)}}$}
    \caption{Distributed price adjustment process.}
    \label{alg:GreedyAlg}
\end{algorithm}%

In Algorithm \ref{alg:GreedyAlg}, the price adjustment process is described. This process is essentially a bisection method which finds the roots of the excess demand function described in the proof of Lemma \ref{thm:uniqueness}. The accuracy measure conditioning the termination of the algorithm is defined as $\epsilon$. The terms $\underline{\beta}$ and $\overline{\beta}$ are the lower and upper bounds on the price ratio given in \eqref{eq:condPrices}, respectively. The prices ratio is initialized to the middle value of these bounds and forwarded to the links. The links send their demands calculated from Theorem \ref{thm:demand} to the arbitrator. If the demand of good $1$ is greater than its supply, then the arbitrator increases the ratio of the prices to half the distance to the upper bound $\overline{\beta}$. Thus, the price of good $1$ relative to the price of good $2$ increases. The lower bound on the prices ratio $\underline{\beta}$ is updated to the price ratio of the previous iteration. If the demand of good $1$ is less than its supply, the price ratio is decremented half the distance to the lower bound $\underline{\beta}$. The upper bound $\overline{\beta}$ is set to the prices ratio of the previous iteration. The algorithm terminates when the distance between the updated upper and lower bounds on the prices ratio is below an accuracy measure $\epsilon$.

\begin{figure}[ht]
\centering
\includegraphics[width=\linewidth,clip]{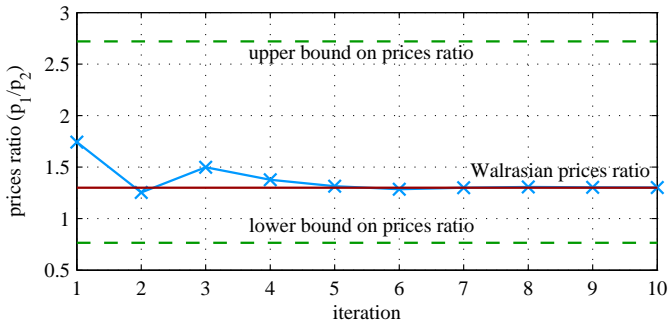}
\caption{\label{fig:tatonnement}Convergence of the price ratio in the price adjustment process to the Walrasian price ratio.}
\end{figure}

In \figurename~\ref{fig:tatonnement}, the prices ratio in the price adjustment process is marked with a cross and is shown to converge after a few iterations to the Walrasian prices ratio from Theorem \ref{thm:Walras}. The dashed lines correspond to the upper and lower bounds in \eqref{eq:condPrices}. 
\section{Discussion and Illustrations}\label{sec:discussion}

\begin{figure}[t]
\centering
\includegraphics[width=\linewidth,clip]{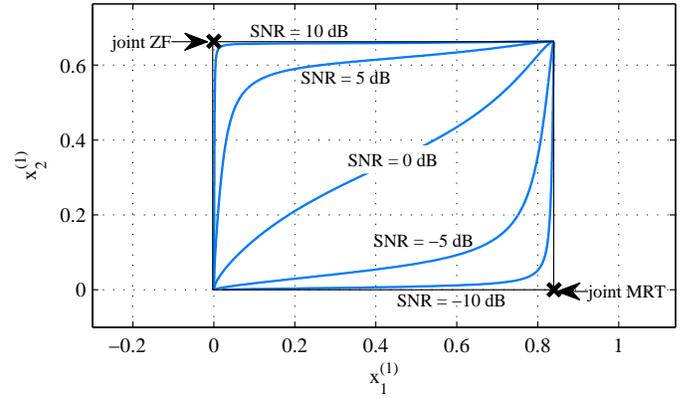}
\caption{\label{fig:contractSNR}Course of the contract curve in the Edgeworth box for different SNR values.}
\end{figure}

In \figurename~\ref{fig:contractSNR}, the contract curve characterized in Theorem \ref{thm:contractcurve} is plotted for different SNR values. The number of antennas at the transmitters is two and we generate independent instantaneous channels $\bh_{k\ell}$ identically distributed as $\mathcal{CN}(0,\bI)$. The contract curve is calculated by taking $10^3$ samples of $\x22$ uniformly spaced in $(0,\lambda_2^\mrt)$ to obtain values of $\x11$. The course of the contract curve for $10$ dB SNR is near to the edge of the Edgeworth box where joint ZF is marked. This means that Pareto optimal allocations require either transmitter to choose beamforming vectors near to ZF. For decreasing SNR, the contract curve moves away from the ZF edge. For low SNR, the contract curve is then close to the edge with joint MRT. These observations conform with the analysis in \cite{Larsson2008a} where Pareto optimal maximum sum utility transmission is studied in low and high SNR regimes.

\begin{figure}[h]
\centering
\includegraphics[width=\linewidth,clip]{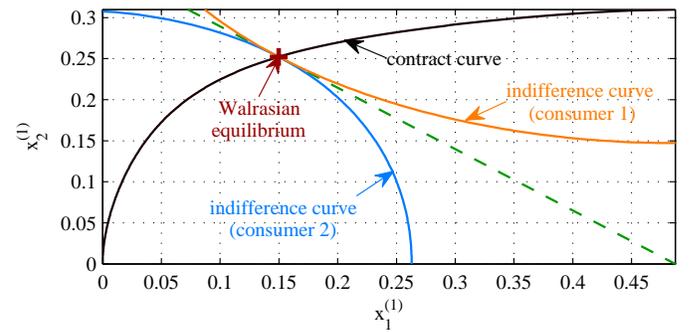}
\caption{\label{fig:edge}Edgeworth box which depicts the allocation for the Walrasian prices.}
\end{figure}

In \figurename~\ref{fig:edge}, an Edgeworth box is plotted for a sample channel realization with two transmit antennas at both transmitters. For the prices calculated from Theorem \ref{thm:Walras} we obtain the Walrasian equilibrium allocation on the contract curve where the corresponding indifference curves are tangent. The indifference curves are obtained from Proposition \ref{thm:ICurve}. The line passing through Walrasian equilibrium allocation defines the budget sets of the consumers as is illustrated in \figurename~\ref{fig:EdgeworthBox_exp}.

\begin{figure}[h]
\centering
\includegraphics[width=\linewidth,clip]{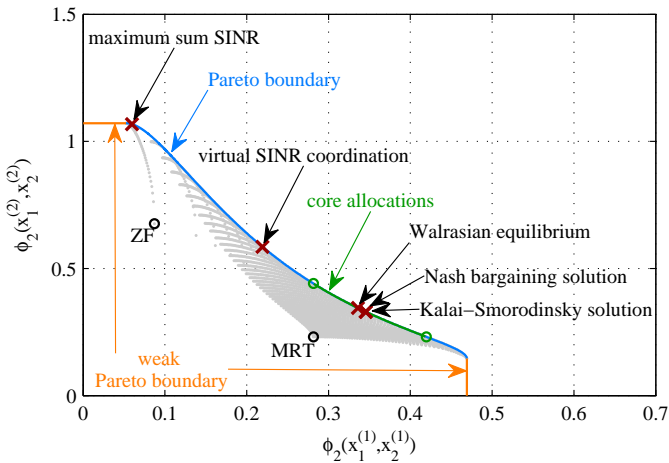}
\caption{\label{fig:region}SINR region of a two-user MISO IFC with SNR = 0 dB and two antennas at the transmitters.}
\end{figure}

In \figurename~\ref{fig:region}, the SINR region is plotted. The points lying inside the SINR region correspond to the beamforming vectors characterized in \eqref{eq:beam_opt_param}, where a subset of these points are Pareto optimal. The Pareto boundary corresponds to the allocations on the contract curve calculated in Theorem \ref{thm:contractcurve}. The weak Pareto boundary consists of weak Pareto optimal points in which the links cannot strictly increase their utility simultaneously. Formally, the weak Pareto optimal points of the SINR region $\Phi$ are defined as \cite[p. 14]{Peters1992}
\begin{equation}\label{eq:WParetoBoundary}
\mathcal{W}(\Phi) :=\br{\mat{x} \in \Phi : \text{there is no } \mat{y} \in \Phi \text{ with } \mat{y} > \mat{x}},
\end{equation}
\noindent where the inequality in \eqref{eq:WParetoBoundary} is componentwise. Pareto optimal points $\mathcal{P}(\Phi)$ in \eqref{eq:ParetoBoundary} define a stronger optimality for a utility tuple than weak Pareto optimal points. A weak Pareto optimal point is not necessarily Pareto optimal. But all Pareto optimal points are also weak Pareto optimal, i.e. $\mathcal{P}(\Phi) \subseteq \mathcal{W}(\Phi)$.

The core allocations are all Pareto optimal points that dominate the Nash equilibrium (joint MRT). Assuming the links are rational, only allocations in the core can be of interest for the links. In other words, the links will not cooperate if one link would achieve lower payoff than at the Nash equilibrium. The Walrasian equilibrium from Theorem \ref{thm:Walras} always lies in the core. In \figurename~\ref{fig:region}, we also plot the maximum sum SINR which is obtained by grid search over the allocations on the Pareto boundary. The virtual SINR coordination point corresponds to the coordination mechanism in \cite{Zakhour2009}, where the minimum mean square error (MMSE) transmit beamforming vectors
\begin{equation}\label{eq:mmse}
\bw_k^\mmse = \frac{[\sigma^2 \bI + \bh_{k\ell} \bh_{k\ell}^H]^{-1} \bh_{kk}}{\norm{[\sigma^2 \bI + \bh_{k\ell} \bh_{k\ell}^H]^{-1} \bh_{kk}}}, \quad k \neq \ell,
\end{equation}
\noindent are proven to achieve a Pareto optimal point. These beamforming vectors require only local channel state information at the transmitters which is an appealing property in terms of the low overhead in information exchange between the links. The virtual SINR coordination and the maximum sum SINR points do not necessarily lie in the core. Hence, these points are not suitable for distributed implementation between the rational links.

In \figurename~\ref{fig:region}, two solutions from axiomatic bargaining theory, namely the Nash bargaining solution (NBS) and the Kalai-Smorodinsky (KS) solution are plotted. These solutions lie in the core and differ by the axioms that define them. The interested reader is referred to \cite{Peters1992} for a comprehensive theory on axiomatic bargaining. According to simulations, these two solutions are not far from each other. The properties that the Walrasian equilibrium and the NBS or KS solution have in common is that they are Pareto optimal and lie in the core, i.e., each user achieves higher utility than at the Nash equilibrium. The difference between the solutions is the fairness aspects in allocating the Pareto optimal utilities to the players. The current advantage in the Walrasian equilibrium over NBS and KS solution is that it can be characterized in closed-form using Theorem \ref{thm:demand} and Theorem \ref{thm:Walras}. In addition, we devise a coordination mechanism to implement the Walrasian equilibrium in Section \ref{sec:coordination}. Next, we will describe how the NBS and KS solutions are obtained. The NBS \cite[Chapter 15]{Osborne1994} is the solution of the following problem:
\begin{equation}\label{eq:NSB}
\begin{split}
    \text{maximize}& \quad (\phi_1 - \phi_1^\ne)(\phi_2 - \phi_2^\ne)\\
    \text{subject to} &\quad (\phi_1,\phi_2) \in \Phi,
\end{split}
\end{equation}
\noindent where $\phi_k^\ne := \phi_k(\lambda_1^\mrt,\lambda_2^\mrt)$ is the SINR in Nash equilibrium and $\Phi$ is the SINR region in \eqref{eq:RR}. Note that the NBS is defined for convex utility regions only, and the SINR region $\Phi$ in our case is not necessarily convex as is shown in \figurename~\ref{fig:region}. However, solving the optimization problem in \eqref{eq:NSB} by grid search over $10^3$ generated Pareto optimal points from Theorem \ref{thm:contractcurve} gives a single solution which we plot in \figurename~\ref{fig:region}. The KS solution is the solution of the following problem \cite{Nokleby2009}:
\begin{equation}\label{eq:KS}
\begin{split}
    \text{maximize} & \quad \min \pp{\frac{\phi_1 - \phi_1^\ne}{\phi_1^\core - \phi_1^\ne}, \frac{\phi_2 - \phi_2^\ne}{\phi_2^\core - \phi_2^\ne}}\\
    \text{subject to} & \quad (\phi_1,\phi_2) \in \Phi,
\end{split}
\end{equation}
\noindent where $\phi_1^\core$ (analogously $\phi_2^\core$) is the solution of the following problem:
\begin{equation}\label{eq:KSbounds}
\begin{split}
    \text{maximize} & \quad \phi_1\\
    \text{subject to} & \quad (\phi_1,\phi_2^\ne) \in \Phi.
\end{split}
\end{equation}
\noindent The two Pareto optimal points $(\phi_1^\core,\phi_2^\ne)$ and $(\phi_2^\ne,\phi_2^\core)$ are the bounds to the core and are marked with circles on the Pareto boundary in \figurename~\ref{fig:region}. These bounds, as discussed in Section \ref{sec:edgeworth}, can be calculated in the Edgeworth box as the intersection of the contract curve and the indifference curves corresponding to the Nash equilibrium. The KS solution which solves the problem in \eqref{eq:KS} using the core bounds is then found by grid search over the generated Pareto optimal points from Theorem \ref{thm:contractcurve}.

\subsection{Difficulties in the Extension to $K$-User MISO IFC}\label{sec:exten}
While the tools in the paper can be applied to general $K$ consumer and $M$ goods economy as can be found in \cite{Jehle2003,Mas-Colell1995}, the application to the beamforming problem in the MISO IFC can currently be done only for the two-user case. This is mainly because of the structure of the parametrization available for the efficient beamforming vectors in the general case.

Using the parametrization in \eqref{eq:beam_opt_param} for two-users, we have chosen in Section \ref{sec:exchange} the amount of good $1$ for consumer $1$ as $\x{1}{1} = \lambda_1$ and the amount of good $1$ for consumer $2$ as $\x{1}{2} = \lambda_1^\mrt - \lambda_1$. With this relation between the parameters and the goods and due to the structure of the expression in \eqref{eq:beam_opt_param}, the SINR in \eqref{eq:utility} for link $1$ depends only on $\x{1}{1}$ and $\x{2}{1}$ which are the amounts from good $1$ and good $2$ for consumer $1$. This method of defining the goods in terms of the parameters does not carry on for the $K$-user MISO IFC case. We illustrate this drawback based on an example in the $3$-user case. The parametrization for the beamforming vectors are \cite{Mochaourab2011a}
\begin{multline}\label{eq:parametrizationKUSER1}
\bw_1 (\lambda_{11},\lambda_{12},\lambda_{13}) \\ = \bv_{max}\pp{\lambda_{11} \bh_{11}\bh_{11}^H - \lambda_{12} \bh_{12}\bh_{12}^H - \lambda_{13} \bh_{13}\bh_{13}^H},
\end{multline}
\begin{multline}\label{eq:parametrizationKUSER2}
\bw_2 (\lambda_{21},\lambda_{22},\lambda_{23}) \\ = \bv_{max}\pp{- \lambda_{21} \bh_{21}\bh_{21}^H + \lambda_{22} \bh_{22}\bh_{22}^H - \lambda_{23} \bh_{23}\bh_{23}^H},
\end{multline}
\begin{multline}\label{eq:parametrizationKUSER3}
\bw_3 (\lambda_{31},\lambda_{32},\lambda_{33}) \\ = \bv_{max}\pp{-\lambda_{31} \bh_{31}\bh_{31}^H - \lambda_{32} \bh_{32}\bh_{32}^H + \lambda_{33} \bh_{33}\bh_{33}^H},
\end{multline}
\noindent where $\bv_{max}(\mat{Z})$ is the eigenvector that corresponds to the largest eigenvalue of $\mat{Z}$ and $\lambda_{k1} + \lambda_{k2} + \lambda_{k3} = 1, k = 1,2,3$. Note that different real-valued parameterizations are also provided in \cite{Shang2011, Zhang2010, Emil2011} which also lead to the same conclusion in terms of the application of the exchange economy model. We use the parametrization in \cite{Mochaourab2011a} in order to highlight the usage of the different parameters. In \eqref{eq:parametrizationKUSER1}-\eqref{eq:parametrizationKUSER3}, three goods can be directly distinguished each corresponding to the parameters of each transmitter. We can choose the amount of good $1$ (analogously for goods $2$ and $3$) to be divided between the three links as $\x{1}{1} = \lambda_{11}$ for link $1$, $\x{1}{2} = \lambda_{12}$ for link $2$, and $\x{1}{3} = \lambda_{13}$ for link $3$. In order to model this setting as an exchange economy, the utility (SINR) of link $k$ should only depend on the amounts of goods $\x{1}{k}, \x{2}{k}, \x{3}{k}$. However, with the parametrization in \eqref{eq:parametrizationKUSER1}-\eqref{eq:parametrizationKUSER3}, the SINR expression of a link $k$ would depend on all parameters. Hence, in formulating the demand of consumer $k$ as is done in the two-user case in \eqref{prb:demand}, the solution depends also on the demands of the other consumers. In this case, each consumer cannot find his optimal demand of goods independently without knowing what the other consumers demand. Due to this fact, it is currently not possible to find the Walrasian equilibrium in the general $K$-user MISO IFC case.

\section{Conclusions}\label{sec:conclusion}
In this work, we model the interaction between two links in the MISO IFC as an exchange economy. The links are considered as the consumers and the exchanged goods correspond to beamforming vectors. Utilizing the conflict representation in the Edgeworth box, all Pareto optimal points could be characterized in closed form. The equilibria of the considered exchange economy are related to a solution concept from coalitional game theory called the core. These allocations are Pareto optimal and dominate the Nash equilibrium of a strategic game between the links. We propose a coordination mechanism between the links which achieves a Pareto optimal outcome in the core. For this purpose, the situation between the links is modeled as a competitive market where now each consumer is endowed with a budget and can consume the goods at specific prices. The equilibrium in this economy is called Walrasian and corresponds to the prices that equate the demand to the supply of goods. The unique Walrasian prices are calculated and the coordination mechanism is executed by an arbitrator that forwards the prices to the consumers. The consumers then calculate in a decentralized manner their optimal demand corresponding to beamforming vectors that achieve the Walrasian equilibrium. This outcome is Pareto optimal and dominates the Nash equilibrium in the SINR region. 
\appendices
\section{Proof of Lemma \ref{lem:powergains}}\label{proof:powergains}
The direct and interference power gains, $\sabs{\bh_{kk}^\H \bw_k(\lambda_k)}$ and $\sabs{\bh_{k \ell}^\H \bw_k(\lambda_k)}, k \neq \ell,$ are calculated as functions of the parameters $\lambda_k$ by using the expression for the beamforming vectors in \eqref{eq:beam_opt_param}. The direct power gain is calculated as:
\begin{align}\nonumber
& \sabs{\bh_{kk}^\H \bw_k(\lambda_k)} \\
&= \pp{\sqrt{\lambda_k} \frac{\bh_{kk}^\H \mat{\Pi}_{\bh_{k \ell}}\bh_{kk}}{\norm{\mat{\Pi}_{\bh_{k \ell}}\bh_{kk}}} + \sqrt{1-\lambda_k} \frac{ \bh_{kk}^\H \mat{\Pi}_{\bh_{k \ell}}^\perp \bh_{kk}}{\norm{\mat{\Pi}_{\bh_{k \ell}}^\perp \bh_{kk}}}}^2 \\
&= \pp{\sqrt{\lambda_k} {\norm{\mat{\Pi}_{\bh_{k \ell}}\bh_{kk}}} + \sqrt{1-\lambda_k} {\norm{\mat{\Pi}_{\bh_{k \ell}}^\perp \bh_{kk}}}}^2.
\end{align}
The interference power is:
\begin{align}\nonumber
& \sabs{\bh_{k \ell}^\H \bw_k(\lambda_k)} \\
&= \sabsl{\sqrt{\lambda_k} \frac{\bh_{k \ell}^\H \mat{\Pi}_{\bh_{k \ell}}\bh_{kk}}{\norm{\mat{\Pi}_{\bh_{k \ell}}\bh_{kk}}} + \sqrt{1-\lambda_k} \frac{ \bh_{k \ell}^\H \mat{\Pi}_{\bh_{k \ell}}^\perp \bh_{kk}}{\norm{\mat{\Pi}_{\bh_{k \ell}}^\perp \bh_{kk}}}}\\
&= \lambda_k \frac{\sabs{\bh_{k \ell}^\H \mat{\Pi}_{\bh_{k \ell}}\bh_{kk}}}{\snorm{\mat{\Pi}_{\bh_{k \ell}}\bh_{kk}}} =  \lambda_k \snorm{\bh_{k \ell}}.
\end{align}
These expressions lead to \eqref{eq:gain_dir} and \eqref{eq:gain_int} in Lemma \ref{lem:powergains}.
\section{Proof of Theorem \ref{thm:quasiconcave}}\label{proof:quasiconcave}
First, it is easy to see that the SINR expression in \eqref{eq:utility} is continuous. The SINR $\phi_k\pps{\x1k,\x2k}$ is strongly increasing with the goods $\x{1}{k}$ and $\x{2}{k}$ if $\phi_k\pps{\tx1k,\tx2k} > \phi_k\pps{\x1k,\x2k}$ whenever $\pps{\tx1k,\tx2k} \neq \pps{\x1k,\x2k}$ and $\pps{\tx1k,\tx2k} \geq \pps{\x1k,\x2k}$ \cite[Definition A1.17]{Jehle2003}. Define the directional derivative of $\phi_k$ at $\pps{\x1k,\x2k}$ in direction $\mat{z}$ as
\begin{multline}\label{eq:dirDeriv0}
\nabla_{\mat{z}} \phi_k\pp{\x1k,\x2k} \\ = \lim_{t\rightarrow 0} \frac{\phi_k\pp{\pp{\x1k,\x2k} + t\mat{z}} - \phi_k\pp{\x1k,\x2k}}{t},
\end{multline}
\noindent Since $\phi_k\pps{\x1k,\x2k}$ is differentiable, the limit above can be given as \cite[Chapter A.2]{Jehle2003}
\begin{equation}\label{eq:dirDeriv}
\nabla_{\mat{z}} \phi_k\pp{\x1k,\x2k} = \nabla \phi_k\pp{\x1k,\x2k} \mat{z},
\end{equation}
\noindent where $\nabla \phi_k\pps{\x1k,\x2k}$ is the gradient of $\phi_k$ at $\pps{\x1k,\x2k}$ written as
\begin{equation}
\nabla \phi_k\pp{\x1k,\x2k} = \pp{\frac{\partial \phi_k\pp{\x1k,\x2k}}{\partial \x{k}{k}},\frac{\partial \phi_k\pp{\x1k,\x2k}}{\partial \x{\ell}{k}}},
\end{equation}
with $\ell \neq k$. The directional derivative of $\phi_k\pps{\x1k,\x2k}$ defines the slope of the tangent to $\phi_k\pps{\x1k,\x2k}$ at the point $\pps{\x1k,\x2k}$ in the direction $\mat{z}$. Hence, if the  directional derivative is positive for $\mat{z} = (z_1 , z_2)^T$ with $z_1$ and $z_2$ nonnegative and satisfying $\norm{\mat{z}} = \sqrt{z_1^2 + z_2^2} = 1$, then the utility function $\phi_k\pps{\x1k,\x2k}$ is strongly increasing. Consequently, the directional derivative in \eqref{eq:dirDeriv} is strictly positive if the components of the gradient $\nabla \phi_k\pps{\x1k,\x2k}$ are strictly positive. The first component of $\nabla \phi_k\pps{\x1k,\x2k}$ is
\begin{multline}\label{eq:grad1}
\frac{\partial \phi_k\pp{\x1k,\x2k}}{\partial \x{k}{k}} \\ = \frac{\pp{\sqrt{\x{k}{k} g_{k}} + \sqrt{(1-\x{k}{k})\gpt{k}}}\pp{\sqrt{ \frac{g_{k}}{\x{k}{k}}} - \sqrt{\frac{\gpt{k}}{1-\x{k}{k}}}}}{\sigma^2 + \lambda_\ell^\mrt \g{\ell}{k} - \x{\ell}{k} \g{\ell}{k}}.
\end{multline}
The partial derivative in \eqref{eq:grad1} is strictly larger than zero when $\x{k}{k} < {g_{k}}/\pp{\gpt{k} + g_{k}}$. Substituting $\gpt{k}$ and $g_{k}$ from Lemma \ref{lem:powergains} we get
\begin{equation}
\x{k}{k}  < \frac{g_{k}}{\gpt{k} + g_{k}} = \frac{\norm{\mat{\Pi}_{\bh_{k \ell}}\bh_{kk}}^2}{\snorm{\bh_{kk}}} = \lambda_k^\mrt.
\end{equation}
Since $\x{k}{k}\in [0,\lambda_k^\mrt]$, the partial derivative in \eqref{eq:grad1} is strictly larger than zero except for $\x{k}{k} = \lambda_k^\mrt$. The second component of $\nabla \phi_k\pps{\x1k,\x2k}$ is
\begin{equation}
\frac{\partial \phi_k\pp{\x1k,\x2k}}{\partial \x{\ell}k} = \g{\ell}{k} \frac{\pp{\sqrt{\x{k}{k} g_{k}} + \sqrt{(1-\x{k}{k})\gpt{k}}}^2}{\pp{\sigma^2 + \lambda_\ell^\mrt \g{\ell}{k} - \x{\ell}{k} \g{\ell}{k}}^2},
\end{equation}
\noindent with $\ell \neq k$, which is strictly larger than zero for $\x{\ell}{k} \in [0,\lambda_\ell^\mrt]$. Hence, the directional derivative in \eqref{eq:dirDeriv} is strictly positive for $\pps{\x1k,\x2k} \in [0,\lambda_1^\mrt] \times [0,\lambda_2^\mrt]$ except for the case $\x{k}{k} = \lambda_k^\mrt$ and $\mat{z} = (1,0)$. Since $\lambda_k^\mrt$ is the upper bound on $\x{k}{k}$, the slope of the function $\phi_k\pps{\x1k,\x2k}$ in the direction $\x{k}{k}$ as is restricted by the condition $\mat{z} = (1,0)$ is not of interest.

Next, we will prove that the SINR function is jointly quasiconcave with the goods. Consider the SINR expression in \eqref{eq:utility}, and define $f(\x{k}{k}) := \pp{\sqrt{\x{k}{k} g_{k}} + \sqrt{(1-\x{k}{k})\gpt{k}}}^2$ and $g(\x{\ell}{k}):={\sigma^2 + \lambda_\ell^\mrt \g{\ell}{k} - \x{\ell}{k} \g{\ell}{k}}$. The function $\phi_k\pps{\x1k,\x2k} = f(\x{k}{k})/ g(\x{\ell}{k})$ is strictly quasiconcave if $f(\x{k}{k})$ is strictly concave and $g(\x{\ell}{k})$ is convex \cite[Proposition 2]{Schaible1983}. It is clear that $g(\x{\ell}{k})$ is convex since the function is linear in $\x{\ell}{k}$. In order to show that $f(\x{k}{k})$ is strictly concave, we build the second derivative of $f(\x{k}{k})$ as follows:
\begin{align}\nonumber
\frac{\text{d}^2 f(\x{k}{k})}{\text{d}^2 \x{k}{k} } &=
\pp{\sqrt{g_k/\x{k}{k}} - \sqrt{ \gpt{k}/(1 - \x{k}{k})}}^2 \\ \nonumber
& \quad - \pp{\sqrt{\x{k}{k} g_k} + \sqrt{(1 - \x{k}{k}) \gpt{k}}} \\
& \quad \times \pp{\sqrt{\frac{g_k}{(\x{k}{k})^3}} + \sqrt{\frac{\gpt{k}}{(1 - \x{k}{k})^3}}} \end{align}
\begin{align}\nonumber
& =  \frac{g_k}{\x{k}{k}} + \frac{\gpt{k}}{(1 - \x{k}{k})} - 2\sqrt{\frac{g_k \gpt{k}}{(1 - \x{k}{k})(\x{k}{k})}} - \frac{g_k}{\x{k}{k}} \\
& \quad - \frac{\gpt{k}}{(1 - \x{k}{k})} - {\sqrt{\frac{(1-\x{k}{k})g_k \gpt{k}}{(\x{k}{k})^3}}} - {\sqrt{\frac{\x{k}{k} g_k \gpt{k}}{(1-\x{k}{k})^3}}}\\ \nonumber
& = - 2\sqrt{\frac{g_k \gpt{k}}{(1 - \x{k}{k})(\x{k}{k})}} - \sqrt{\frac{(1-\x{k}{k})g_k \gpt{k}}{(\x{k}{k})^3}} \\
&\quad - \sqrt{\frac{\x{k}{k} g_k \gpt{k}}{(1-\x{k}{k})^3}} < 0.
\end{align}
\noindent The second derivative of $f(\x{k}{k})$ is strictly less than zero. Thus, $f(\x{k}{k})$ is strictly concave. Accordingly, $\phi_k\pps{\x1k,\x2k}$ is strictly quasiconcave.

\section{Proof of Theorem \ref{thm:demand}}\label{proof:demand}
Since the function $\phi_k(\x1k,\x2k)$ is strictly quasiconcave, then this function has a unique maximum. Considering consumer $1$ (analogously consumer $2$), the Lagrangian function to the constrained optimization problem in \eqref{prb:demand} is
\begin{equation}
\begin{split}
\mathcal{L}\pp{\x11,\x21,\mu} & = \phi_1\pp{\x11,\x21}\\
& \quad + \mu \pp{\lambda_1^\mrt p_1 - \x11 p_1 - \x21 p_2},
\end{split}
\end{equation}
\noindent where $\mu$ is a Lagrange multiplier. The Karush--Kuhn--Tucker (KKT) conditions for optimality are necessary and sufficient given as:
\begin{align}\label{ex:KKT11}
\frac{\partial \mathcal{L}\pp{\x11,\x21,\mu}}{\partial \x11} & = \frac{\partial \phi_1\pp{\x11,\x21}}{\partial \x11} - \mu p_1 = 0\\ \label{ex:KKT12}
\frac{\partial \mathcal{L}\pp{\x11,\x21,\mu}}{\partial \x21} & = \frac{\partial \phi_1\pp{\x11,\x21}}{\partial \x21} + \mu p_2 = 0\\ \label{ex:KKT13}
\frac{\partial \mathcal{L}\pp{\x11,\x21,\mu}}{\partial \mu} & = \lambda_1^\mrt p_1 - \x11 p_1 - \x21 p_2 = 0
\end{align}
\noindent According to conditions \eqref{ex:KKT11} and \eqref{ex:KKT12}, we get
\begin{equation}\label{eq:KKTeq1eq2}
\frac{\partial \phi_1\pp{\x11,\x21}}{\partial \x11} \frac{1}{p_1}  = - \frac{\partial \phi_1\pp{\x11,\x21}}{\partial \x21} \frac{1}{p_2}
\end{equation}
\begin{multline}
\Rightarrow\frac{\pp{\sqrt{\x11 g_{1}} + \sqrt{(1-\x11)\gpt{1}}} \pp{\frac{\sqrt{g_1}}{\sqrt{\x11}} - \frac{\sqrt{\zfg_1}}{\sqrt{1-\x11}}}}{\sigma^2 + \lambda_2^\mrt \g{2}{1} - \x21 \g{2}{1}} \\ = \frac{\pp{\sqrt{\x11 g_{1}} + \sqrt{(1-\x11)\gpt{1}}}^2 g_{21} }{\pp{\sigma^2 + \lambda_2^\mrt \g{2}{1} - \x21 \g{2}{1}}^2} \frac{p_1}{p_2}
\end{multline}
\begin{multline}
\Rightarrow{\frac{\sqrt{g_1}}{\sqrt{\x11}} - \frac{\sqrt{\zfg_1}}{\sqrt{1-\x11}}} \\ = \frac{\pp{\sqrt{\x11 g_{1}} + \sqrt{(1-\x11)\gpt{1}}} g_{21} }{(\sigma^2 + \lambda_2^\mrt \g{2}{1} - \x21 \g{2}{1})} \frac{p_1}{p_2}.
\end{multline}
\noindent Substituting $\x21$ from \eqref{ex:KKT13} we get
\begin{multline}\label{eq:B}
{{\sqrt{(1-\x11)g_1}} - \sqrt{\x11 \zfg_1}} \\ = \frac{\pp{\x11 \sqrt{(1-\x11) g_{1}} + (1-\x11) \sqrt{\x11\gpt{1}}}  }{(\underbrace{\frac{\sigma^2}{g_{21}} + \lambda_2^\mrt - \lambda_1^\mrt \frac{p_1}{p_2}}_{B} + \x11 \frac{p_1}{p_2})} \frac{p_1}{p_2}
\end{multline}
\begin{multline}
\Rightarrow{\sqrt{(1-\x11)g_1}B - \sqrt{\x11 \zfg_1}B} - \x11 \frac{p_1}{p_2} \sqrt{\x11 \zfg_1} \\= (1-\x11) \sqrt{\x11\gpt{1}} \frac{p_1}{p_2}
\end{multline}
\begin{equation}
\Rightarrow \sqrt{\x11 \zfg_1}\pp{B + \frac{p_1}{p_2}} = \sqrt{(1-\x11)g_1}B
\end{equation}
\noindent Squaring both sides on the condition that $B \geq 0$ we can write
\begin{align}
{\x11 \zfg_1}\pp{B + \frac{p_1}{p_2}}^2 &= {(1-\x11)g_1}B^2.
\end{align}
\noindent We solve for $\x11$ to get
\begin{align}
{\x11} &= \pp{1 + \frac{\zfg_1}{g_1}\pp{1+\frac{p_1}{p_2 B}}^2}^{-1}.
\end{align}
\noindent Substituting $B$ from \eqref{eq:B} we get the expression in \eqref{eq:demand1}. $\x21$ is calculated according to \eqref{ex:KKT13}.

\bibliographystyle{IEEEtran}
\bibliography{references}
\end{document}